\documentclass[usenatbib]{mnras}

\usepackage{graphicx}
\usepackage{amsmath}	
\usepackage[dvipsnames]{xcolor}  
\usepackage{amssymb}
\usepackage{newtxtext,newtxmath}

\usepackage{ulem}

\newcommand{\ie}{i.e.,~}
\newcommand{\eg}{e.g.,~}

\begin{document}

\title[Non-linear evolutions of magnetised thick discs around black holes: dependence on the initial data]
{Non-linear evolutions of magnetised thick discs around black holes: dependence on the initial data}

\author[A.~Cruz-Osorio, S.~Gimeno-Soler, and J.A.~Font]{Alejandro Cruz-Osorio$^{1}$, Sergio Gimeno-Soler$^{1}$ and Jos\'e A.~Font$^{1,2}$\\ 
$^{1}$Departamento de Astronom\'ia y Astrof\'isica, Universitat de Val\`encia, C/ Dr.  Moliner 50, 
46100, Burjassot (Val\`encia), Spain\\
$^{2}$Observatori Astron\`omic, Universitat de Val\`encia, C/ Catedr\'atico 
  Jos\'e Beltr\'an 2, 46980, Paterna (Val\`encia), Spain}

\maketitle
\begin{abstract}
We build equilibrium solutions of magnetised thick discs around a highly spinning Kerr black hole and evolve these initial data up to a final time of about 100 orbital periods. The numerical simulations reported in this paper solve the general relativistic magnetohydrodynamics equations using the {\tt BHAC} code and are performed in axisymmetry. Our study assumes non-self-gravitating, polytropic, constant angular momentum discs endowed with a purely toroidal magnetic field. In order to build the initial data we consider three approaches, two of which incorporate the magnetic field in a self-consistent way and a third approach in which the magnetic field is included as a perturbation on to an otherwise purely hydrodynamical solution. To test the dependence of the evolution on the initial data, we explore four representative values of the magnetisation parameter spanning from almost hydrodynamical discs to very strongly magnetised tori. The initial data are perturbed to allow for mass and angular momentum accretion on to the black hole. Notable differences are found in the long-term evolutions of the initial data. In particular, our study reveals that highly magnetised discs are unstable, and hence prone to be fully accreted and expelled, unless the magnetic field is incorporated into the initial data in a self-consistent way.
\end{abstract}


\section{Introduction}
\label{sec:intro}

Astrophysical systems consisting of stellar mass black holes surrounded by thick discs (or tori) are broadly regarded as natural end results of catastrophic events involving compact objects. To a significant extent, our theoretical understanding of the formation of those systems has been built from ever more accurate numerical simulations. Two distinctive examples that keep receiving major numerical attention are binary mergers formed by either two neutron stars or by a black hole and a neutron star. Numerical work has shown that those types of mergers may quite generically lead to rotating black holes surrounded by geometrically thick accretion discs (see, e.g.~\cite{ShibataTaniguchilrr-2011-6,Baiotti2016} and references therein). 

Likewise, understanding the long-term dynamics of black hole-torus systems also requires to perform time-dependent numerical simulations. Most studies have made use of a rather simplistic model in which the specific angular momentum of the disc is assumed to be constant. In a purely hydrodynamical context this model is commonly refereed to as a `Polish doughnut', after the seminal work by~\cite{Abramowicz78} (but see also~\cite{Fishbone76}). The extension to the MHD regime of a constant angular momentum disc endowed with a toroidal magnetic field was achieved by~\cite{Komissarov2006a} (see also~\cite{Gimeno-Soler:2017} for the non-constant angular momentum case and~\cite{Pimentel2018a, Pimentel2018b} for models including magnetic polarisation). Polish doughnuts have been extensively used to study instabilities of accretion flows onto black holes (e.g.~the runaway instability~\citep{Abramowicz83} and the Papaloizou-Pringle instability (PPI)~\citep{Papaloizou84}) and the formation of jets and outflows (see e.g.~\cite{Font02a,DeVilliers03,Daigne04,Fragile:2007dk,Dexter2011,Dexter2012,McKinney2012, McKinney2014,Wielgus2015, Fragile2017, Bugli2018, Witzany2018,Janiuk2018}). In all of these works the self-gravity of the fluid/MHD is neglected in the construction of the equilibrium configurations and in the subsequent time evolutions. Equilibrium solutions of self-gravitating tori around black holes, for which the initial data satisfy the constraint equations of the coupled Euler-Einstein system, have been obtained in the purely hydrodynamical constant angular momentum case by~\cite{Shibata2007} (see also~\cite{Mach:2019} for the magnetised non-constant angular momentum case) and by~\cite{Stergioulas2011} (see~\cite{Korobkin2011,Mewes2016} for numerical relativity simulations of those solutions). Moreover, \cite{Shibata2012} obtained solutions of self-gravitating and magnetised tori accounting for the coupled system of radiation, general relativistic MHD and the Einstein equations.

The way the magnetic field is accounted for in the equilibrium solutions is, for most approaches in the literature, essentially arbitrary, \ie its influence on the disc morphology is not treated in a self-consistent fashion. As a result, the initial distribution and strength of the magnetic field in the torus may impact the subsequent time evolution and lead to potential inaccuracies. Early attempts, e.g.~\cite{Koide99}, were based on equilibrium hydrodynamical solutions of a disc around a black hole which was arbitrarily seeded by a uniform magnetic field in the direction perpendicular to the disc. In most recent approaches, the magnetic field distribution is derived from an `ad hoc' guess for the vector potential. This allows to study both poloidal and toroidal configurations of the magnetic field and sets the framework to study the growth of the magneto-rotational instability (MRI), the redistribution of the angular momentum and the accretion mechanism itself~\citep{DeVilliers03a,DeVilliers03,Gammie03,Anninos05c,Noble2006,McKinney2009,Hawley2011,McKinney2012,Shiokawa2012,Sorathia2013,Penna2013b,Foucart2015b,Anninos2017,Porth2017,Mizuta2018}. Similar configurations have been used in recent MHD simulations in general relativity of mini-discs in binary black hole mergers~\citep{Bowen2018}, neutrino-cooled thick accretion discs~\citep{Siegel2017,Siegel2018}, or to compute the shadows around the black holes of Sgr$A^{*}$~\citep{Chan15} or M87$^{*}$~\citep{EHTI,EHTVI}.

In this paper we study whether the way the initial magnetic field distribution in a thick disc is built has an impact on the long-term dynamics of the system and, if so, how significant. 

To this aim we build magnetised Polish doughnuts around rotating black holes, neglecting the self-gravity of the discs and using three different approaches to account for the magnetic field, namely:  i) a purely hydrodynamical solution (see e.g.~\cite{Abramowicz78,Font02a,Daigne04}) in which an `ad hoc' toroidal magnetic field is seeded afterwards;  ii)  the self-consistent solution from~\cite{Komissarov2006a}, in which the distribution of the rest-mass density of the disc is coupled to the toroidal magnetic field through the equation of state for the magnetic pressure; this approach assumes that the fluid is thermodynamically non-relativistic; and iii) the self-consistent approach of~\cite{Komissarov2006a} but dropping the assumption of a thermodynamically non-relativistic fluid, as done in~\citet{Montero07} and~\citet{Gimeno-Soler:2019}. Using these three approaches we build initial data and compare their non-linear dynamical evolutions by means of axisymmetric numerical simulations, finding interesting differences. Our study has been limited to axisymmetry to reduce the computational cost involved in the simulations, since we are interested in the long-term dynamics of the discs, which are evolved up to 100 orbital periods. We note that the first approach has been employed in some general relativistic magneto-hydrodynamics (GRMHD) simulations of magnetised thick discs (e.g.~\citet{Gammie03,Noble2006,Shiokawa2012,Porth2017,Mizuta2018,Bowen2018}) albeit for poloidal configurations of the magnetic field which are MRI unstable. 

All configurations considered in this paper are purely toroidal. Currently, our self-consistent approach to build stationary magnetised discs around black holes can only accommodate toroidal magnetic fields. We plan to extend our approach to poloidal magnetic fields in the future and, if possible, perform a similar comparison with the ad hoc poloidal magnetic field configurations employed in the literature.

The paper is organised as follows: In Section~\ref{sec:Num} we summarise the problem setup, i.e.~the equations of general relativistic MHD and the numerical code. Section~\ref{sec:InitTori} describes the three types of approaches we follow to construct the initial data for magnetised tori. The results of the time evolutions and the comparison among the three approaches are presented in Section~\ref{sec:Results}. Finally Section~\ref{sec:Sum} summarises our conclusions. Unless stated otherwise we use geometrised units in which the light speed, Newton's constant, and the mass of the black hole are equal to one, $c=G=M=1$, the Kerr metric has the signature $(-,+,+,+)$, and the $1/4\pi$ factor in the MHD equations is assumed to be one.     

\section{Setup}
\label{sec:Num}

To describe the Kerr black hole spacetime we use horizon-penetrating Kerr-Schild coordinates with a logarithmic radial coordinate. In the 3+1 decomposition the line element and metric potentials are written as
\begin{eqnarray}
ds^{2}&=& -(\alpha^{2} - \beta_{i}\beta^{i})dt^{2} + 2\beta_{i}dx^{i}dt + \gamma_{ij}dx^{i}dx^{j}, \label{eq:line}\\ 
\alpha &=& \left( 1+{2Me^{R}}/{\varrho^2}\right)^{-1/2}, \nonumber \\
\beta^R &=& e^{R}\frac{2M}{\varrho^2}\left( 1+{2Me^{R}}/{\varrho^2}\right)^{-1}, \nonumber \\
\gamma_{RR} &=& \left( 1+{2Me^{R}}/{\varrho^2}\right)e^{2R} ,~~~~~\gamma_{\theta\theta} = \varrho^2, \nonumber \\
\gamma_{R\phi} &=&  -ae^{R}\left(1+{2Me^{R}}/{\varrho^2}\right)\sin^2\theta,\nonumber \\
\gamma_{\phi\phi} &=& \sin^2\theta ~[~ \varrho^2+a^{2}\left(1+{2Me^{R}}/{\varrho^2}\right)\sin^2\theta ]\,\nonumber 
\end{eqnarray}
where $M$ stands for mass of the black hole and $a = J/M$ is the rescaled angular momentum of the black hole. Note that, in the above expressions the lapse function $\alpha$, the shift vector $\beta^i$ and the three-metric components $\gamma_{ij}$ are written using a modified Kerr-Schild coordinate such as $r=e^{R}$ (and then $\varrho^2 \equiv e^{2R} + a^2\cos^2\theta$).

The general relativistic ideal MHD (GRMHD) evolution equations are obtained from the baryon number conservation, the local conservation of the energy-momentum tensor $T^{\mu\nu}$ and the Maxwell equations
\begin{eqnarray}
\nabla_{\mu} (\rho u^{\mu}) &=& 0 \,, \\
\nabla_{\mu} T^{{\mu \nu}} &=& 0 \,, \label{eq:conservation}\\
\nabla_{\mu}\, ^{*}\!F^{\mu\nu} &=& 0 \,, 
\label{eq:grmhd}
\end{eqnarray}
where $\rho$ is the rest-mass density, and $F^{\mu\nu}$ and $^{*}\!F^{\mu\nu}=b^\mu u^\nu - b^\nu u^\mu$ are the Faraday tensor and its dual with respect to an observer with four-velocity $u^{\mu}$, respectively. The energy-momentum tensor for a magnetised perfect fluid can be written as
\begin{eqnarray}
\label{eq:energy-momentum_tensor}
T^{\mu\nu} = \rho h_{\rm tot}u^{\mu}u^{\nu} + p_{\rm tot} g^{\mu \nu} -b^{\mu}b^{\nu}, 
\end{eqnarray}
where $h_{\rm tot}= 1 + \epsilon + p/\rho + b^{2}/\rho$ is the total specific enthalpy, $p_{\rm tot} = p + p_{\rm m}$ is the total pressure and $p_{\rm m} = b^2/2$ can be seen as the magnetic field contribution to the total pressure, and $b^{2}=b_{\mu}b^{\mu}$ is the square of the magnetic field four-vector. Given the spacetime metric we can write the GRMHD equations in flux-conservative form, in the so-called Valencia formulation \citep[for  details see][]{Anton05, Porth2017}.

\begin{figure*}
\includegraphics[width=0.85\columnwidth]{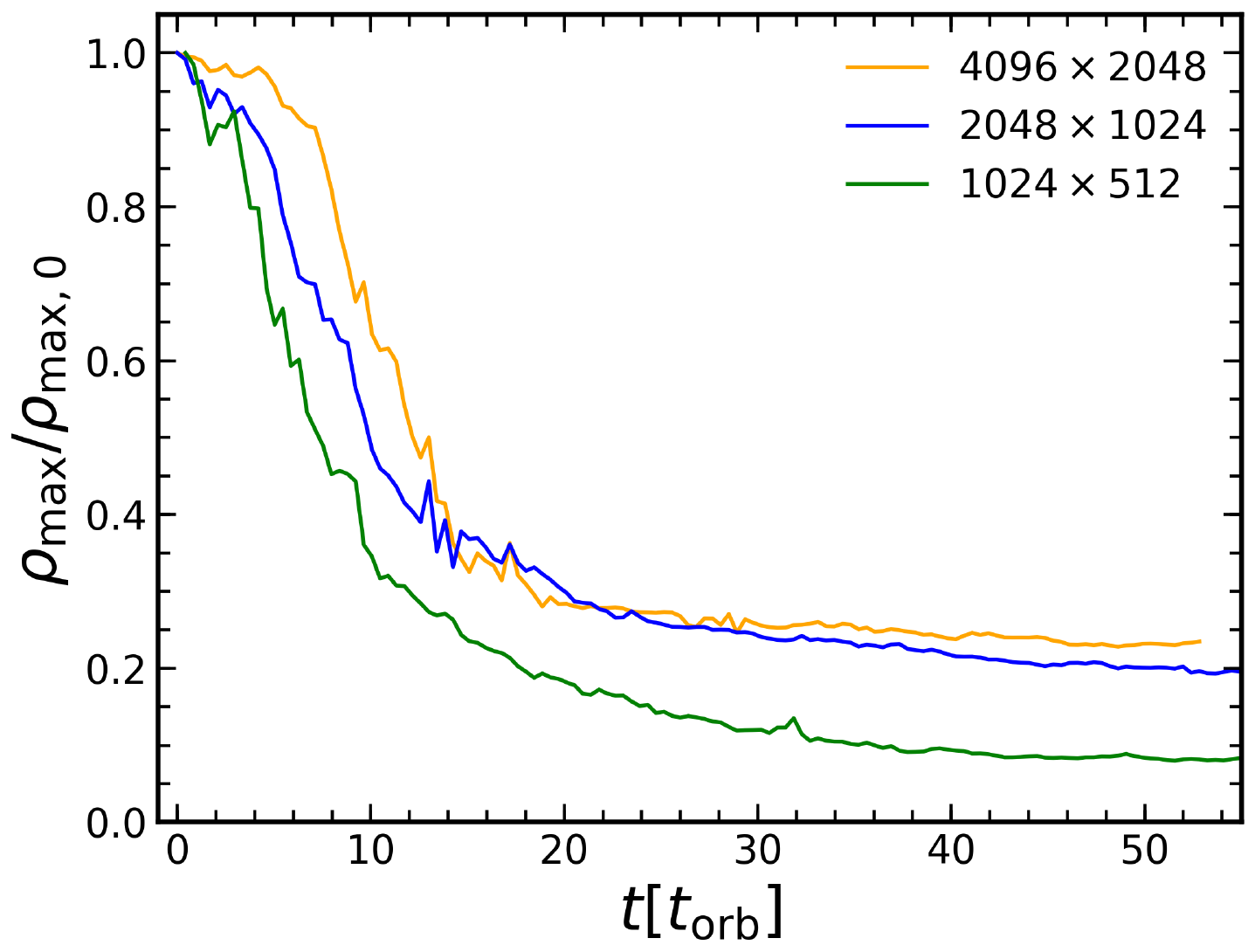}
\hspace{0.3cm}
\includegraphics[width=0.85\columnwidth]{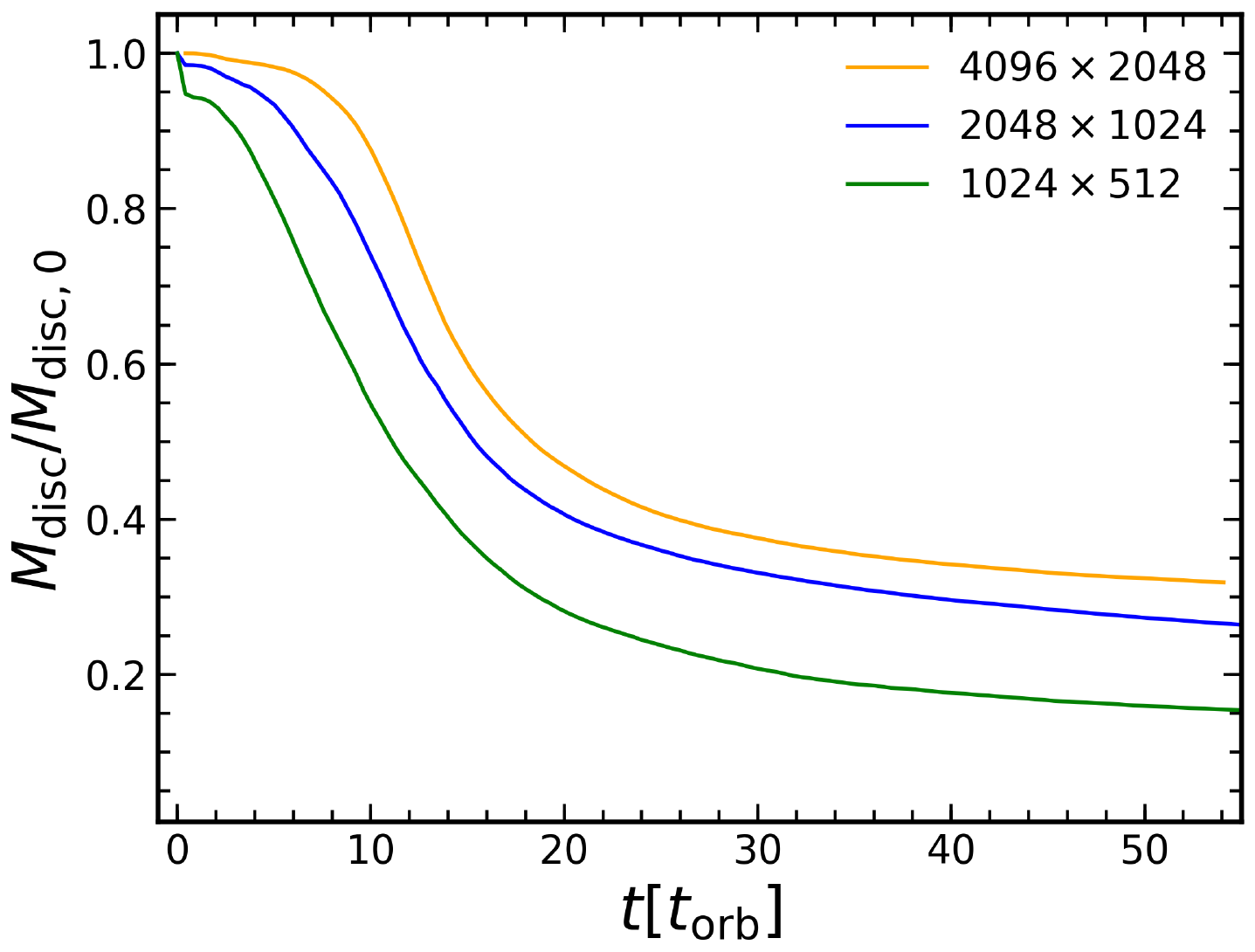}
\caption{ Grid resolution comparison: evolution of the maximum rest-mass density normalised by the initial value (left) and of the mass of the disc (right) for model \texttt{KM}$\rho$-D. The time is given in units of the orbital period at the centre of the disc and the effective number of zones employed in each simulation are indicated in the legend. For all grid resolutions,  the density maximum and the mass of the disc gradually decrease as a result of accretion until they reach constant asymptotic values. With the standard base grid used in our simulations, the final values of the two quantities are underestimated by about 10\%.}
\label{fig:resolution}
\end{figure*}

The numerical simulations reported in this paper are performed in axisymmetry using the \texttt{BHAC} code~\citep{Porth2017}. This code solves the GRMHD equations with a third-order Runge-Kutta method of lines~\citep{Shu88} together with high-resolution shock-capturing algorithms. We use the HLLE two-wave flux formula~\citep{Harten83, Einfeldt88} and a modified five-order WENO-Z cell reconstruction scheme~\citep{Acker2016}. We note that modern extensions of HLLE such as the five-wave HLLD 
 method~\citep{Mignone2009, Matsumoto2019}, where the complete fan of waves is considered, are not yet implemented in \texttt{BHAC}. The preservation of the no magnetic monopoles constriction is achieved by using  the flux constrained transport method~\citep[for more details see][]{Olivares2018b}. Primitive variables are recovered using the inversion technique 2DW from~\cite{Noble2006}. The density and the pressure of the atmosphere outside of the magnetised tori used in our simulations are $\rho_{\rm atm}=\rho_{0} r^{-3/2}$ and $p_{\rm atm}= p_{0} r^{-5/2}$ where $\rho_{0}=10^{-5}$ and $p_{0}=10^{-7}$ as used in~\cite{Noble2006}. In addition, the magnetic field is set to zero if $\rho\le \rho_{\rm atm}$.

Both to build the initial data and in the time-dependent simulations we use a numerical $(r,\theta)$ grid with three refinement levels in a domain $r\in [0.5M,1000M]$. Outflow boundary conditions are applied in the radial direction and reflecting boundary conditions in the angular direction. We use three levels of octree adaptive mesh refinement in the base grid with $512 \times 256$ zones in $r \times \theta$, respectively.  The error estimator formula from~\cite{Loehner87} is applied to the rest-mass density and magnetic field with a tolerance of $0.1$. This error is monitored every 1000 time iterations, changing the grid resolution when necessary. Our two finer grids have thus $1024 \times 512$ and $2048 \times 1024$ zones, respectively. Test runs with a factor 2 coarser and finer grids have been carried out for validation purposes, as displayed in Fig.~\ref{fig:resolution}. This figure shows the time evolution for over 50 orbital periods of the (normalised) mass of one of our accretion discs and of the maximum of the rest-mass density. For this grid comparison we employ a representative highly magnetised model of our sample (namely, case D of model \texttt{KM}$\rho$; see below). We explore three effective resolutions, with $1024 \times 512$, $2048 \times 1024$ and $4096 \times 2048$ cells, respectively. In the two quantities plotted in Fig.~\ref{fig:resolution} we can see that, as a result of accretion on to the black hole, the density maximum and the mass of the disc gradually decrease until they reach constant asymptotic values. While the particular final values are sensitive to the resolution employed, the actual trend is similar for all resolutions. From this figure we conclude that, with our standard base grid, the final values of the maximum density and of the mass of the disc are underestimated by about 10\%. Keeping this in mind, and considering that employing a high-resolution grid with $4096 \times 2048$ zones would be highly time-consuming (even in axisymmetry) for our long-term evolutions extending up to about 100 orbital periods, we use the standard base grid in all results discussed in this paper.

\section{Initial data for magnetised thick discs}
\label{sec:InitTori}

Since the standard procedure to build a stationary accretion disc around a Kerr black hole is well known, we will only sketch it here skipping most details. The interested reader is addressed to~\cite{Abramowicz78,Komissarov2006a,Montero07} for details.

We begin by assuming a stationary and axisymmetric fluid field in a Kerr background. Also, we consider a purely toroidal magnetic field (i.e.~$b^r = b^{\theta} = 0$). By contracting the conservation law for the energy-momentum tensor with the projection tensor $h^{\alpha}_{\,\,\beta} = \delta^{\alpha}_{\,\,\beta} + u^{\alpha}u_{\beta}$ and following~\cite{Komissarov2006a}, we can rewrite the conservation law in terms of the specific angular momentum $l = - u_{\phi}/u_t$ and of the angular velocity $\Omega = u^{\phi}/u^t$, to obtain
\begin{equation}\label{eq:diff_ver}
\partial_i(\ln u_t|) - \frac{\Omega \partial_i l}{1-l\Omega} + \frac{\partial_i p}{\rho h} + \frac{\partial_i(\mathcal{L}b^2)}{2\mathcal{L} \rho h} = 0\,,
\end{equation}
where $i = r, \theta$ and $\mathcal{L} \equiv g_{t\phi}^2 - g_{tt}g_{\phi\phi}$. It is also useful to introduce the definition of total (gravitational plus centrifugal) potential~\citep{Abramowicz78} as
\begin{equation}\label{eq:potential}
W = \ln |u_t| - \int^{l_{\infty}}_{l}\frac{\Omega \partial_i l}{1-l\Omega}.
\end{equation}
From this point on, and as we previously mentioned, we take three different approaches to integrate Eq.~\eqref{eq:diff_ver} which are discussed next.

\subsection{Non-magnetised torus plus toroidal magnetic field}
\label{sub:MFD}

Following the procedure described by~\cite{Font02a} we can construct a non-magnetised torus and subsequently seed it with a magnetic field. We denote the corresponding model as \texttt{MFD}.  To do this, we simply need to take $b = 0$ in Eq.~\eqref{eq:diff_ver}. Then, assuming a constant distribution of angular momentum and a barotropic equation of state (EoS) $\rho = \rho(p)$ we can rewrite Eq.~\eqref{eq:diff_ver} as
\begin{equation}
\mathrm{d}\left(\ln |u_t| + \int^p_0\frac{\mathrm{d}p}{\rho h}\right) = 0\,.
\end{equation}
At the inner edge of the disc we assume $u_t = u_{t_{\mathrm{in}}}$ and $p = 0$, and we can rewrite the above equation as 
\begin{equation}\label{eq:FD_preint}
W - W_{\mathrm{in}} + \int^p_0 \frac{\mathrm{d}p}{\rho h} = 0\,
\end{equation}
where we have used the definition of the potential, Eq.~\eqref{eq:potential}.
Using a polytropic EoS $p = K \rho^{\Gamma}$, with $K$ and $\Gamma$ constants, and the definition of the specific enthalpy, we can integrate Eq.~\eqref{eq:FD_preint}
\begin{equation}
W_{\mathrm{in}} - W = \ln \frac{h}{h_{\mathrm{in}}}\,,
\end{equation}
which can be rewritten as
\begin{equation}\label{eq:FD_final_enthalpy}
h = h_{\mathrm{in}} e^{\Delta W}\,,
\end{equation}
where $\Delta W = W_{\mathrm{in}} - W$.
Then, we can write the expressions for the rest-mass density and the fluid pressure
\begin{equation}\label{eq:FD_density}
\rho = \left(\frac{\Gamma-1}{\Gamma}\frac{(h_{\mathrm{in}} e^{\Delta W} -1)}{K}\right)^{1/(\Gamma-1)}\,,
\end{equation}
\begin{equation}\label{eq:FD_pressure}
p = \left(\frac{\Gamma-1}{\Gamma}\frac{(h_{\mathrm{in}} e^{\Delta W} -1)}{K^{1/\Gamma}}\right)^{\Gamma/(\Gamma-1)}.
\end{equation}

To complete the model, following \cite{Porth2017} we add an `ad hoc' toroidal magnetic field in the following way: we choose a value for the magnetisation parameter $\beta_{\mathrm{m}}=p/p_{\mathrm{m}}$ and insert Eq.~\eqref{eq:FD_pressure} in its definition to arrive at
\begin{equation}\label{eq:FD_pm}
p_{\mathrm{m}} = \frac{1}{\beta_{\mathrm{m}}}\left(\frac{\Gamma-1}{\Gamma}\frac{(h_{\mathrm{in}} e^{\Delta W} -1)}{K^{1/\Gamma}}\right)^{\Gamma/(\Gamma-1)},
\end{equation}
which provides the magnetic pressure in this approach. Note that, in this model, the ratio between the pressure $p$ and the magnetic pressure $p_{\mathrm{m}}$ (\ie $\beta_{\mathrm{m}}$) remains constant throughout the disc. To obtain the non-zero components of the magnetic field, we use
\begin{eqnarray}\label{eq:FD_mfield}
b^{\phi} &=& \sqrt{\frac{2p_{\mathrm{m}}}{{\cal A}}}, \\
b^{t}&=& l b^{\phi}, 
\end{eqnarray} 
where ${\cal A}\equiv g_{\phi \phi} + 2lg_{t\phi} + l^{2}g_{tt}$.

\subsection{Magnetised torus plus relativistic fluid}
\label{sub:KMH}

Our second approach follows the procedure described in~\cite{Montero07} which takes into account the magnetic field from the beginning to construct the disc in a self-consistent way. We denote this model as \texttt{KM$\rho$}. First, we choose a barotropic EoS $\rho = \rho(p)$ of the same form as before
\begin{equation}\label{eq:polytrope}
p = K \rho^\Gamma\,,
\end{equation}
and we introduce the magnetic pressure, $p_{\mathrm{m}} = b^2/2$, and the following quantities: $w = \rho h$, , $\tilde{w} = \mathcal{L} w$ and $\tilde{p}_{\mathrm{m}} = \mathcal{L} p_{\mathrm{m}}$. We can write a similar equation to Eq.~\eqref{eq:polytrope} for the magnetic pressure
\begin{equation}
\tilde{p}_{\mathrm{m}} = K_{\mathrm{m}} \tilde{w}^{q}\,,
\end{equation}
where $K_{m}$ and $q$ are constants. In terms of the magnetic pressure, this equation reads
\begin{equation}\label{eq:mag_pressure}
p_{\mathrm{m}} = K_{\mathrm{m}} \mathcal{L}^{q-1}w^{q}\,.
\end{equation}
This particular choices of EoS for the fluid pressure and the magnetic pressure fulfill the general relativistic version of the von Zeipel theorem for a toroidal magnetic field~\citep{vonZeipel1924, Zanotti2015d}. This allows us to integrate Eq.~\eqref{eq:diff_ver}
\begin{equation}
\ln |u_t| - \int^l_0 \frac{\Omega \mathrm{d}l}{1 - \Omega l} + \int^p_0 \frac{\mathrm{d}p}{\rho h} + \int_0^{\tilde{p}_{\mathrm{m}}} \frac{\mathrm{d}\tilde{p}_{\mathrm{m}}}{\tilde{w}} = \mathrm{const}.
\end{equation}
Following the same reasoning as in the previous section, we can find the constant of integration as
\begin{equation}
\mathrm{const.} = \ln |u_t| - \int^l_{l_\mathrm{in}} \frac{\Omega \mathrm{d}l}{1 - \Omega l}.
\end{equation}
If we insert in this expression the definition of the total potential Eq.~\eqref{eq:potential} we can rewrite the previous expression as
\begin{equation}\label{eq:MK_preint}
W - W_{\mathrm{in}} = \int^p_0 \frac{\mathrm{d}p}{\rho h} + \int_0^{\tilde{p}_{\mathrm{m}}} \frac{\mathrm{d}\tilde{p}_{\mathrm{m}}}{\tilde{w}}.
\end{equation}
Substituting the EoS and taking into account that our fluid is ideal and isentropic, we can integrate Eq.~\eqref{eq:MK_preint} as
\begin{equation}\label{eq:MK_enthalpy}
W - W_{\mathrm{in}} + \ln \left(\frac{h}{h_{\mathrm{in}}}\right) + \frac{q}{q-1}K_{\mathrm{m}}(\mathcal{L}w)^{q-1} = 0\,,
\end{equation}
where we have used that $p_{\mathrm{in}} = p_{\mathrm{m, in}} = \rho_{\mathrm{in}} = 0$. We can rewrite this equation in terms of the rest-mass density $\rho$
\begin{eqnarray}\label{eq:MK:final_eq}
W - W_{\mathrm{in}} + \ln \left(1+ \frac{K\Gamma}{\Gamma-1}\rho^{\Gamma-1}\right) + \nonumber \\ \frac{q}{q-1}K_{\mathrm{m}}\left[\mathcal{L}\left(\rho + \frac{K\Gamma \rho^{\Gamma}}{\Gamma-1}\right)\right]^{q-1} = 0.
\end{eqnarray}
We should note that Eq.~\eqref{eq:MK_enthalpy} is equivalent to Eq.~\eqref{eq:FD_final_enthalpy} in the previous section for a non-magnetised flow ($K_{\mathrm{m}} = 0$). Also, it is important to note that Eq.~\eqref{eq:MK:final_eq} is a trascendental equation and must be solved numerically.

\subsection{Magnetised torus plus non-relativistic fluid}
\label{sub:KMRHO}

We describe next our third procedure to build a magnetised torus. This one is based on the approach introduced by~\cite{Komissarov2006a}. This solution is obtained by assuming the rest-mass density $\rho$ to be almost equal to the fluid enthalpy $\rho \simeq w$ (i.e.~$h \simeq 1$). This approximation means that the fluid is non-relativistic from a thermodynamical point of view. We denote the corresponding disc model as \texttt{KM$h$}.

Since $\rho \simeq w$, we rewrite Eq.~\eqref{eq:polytrope} as
$p = K w^{\Gamma}$.
Substituting this into the definition of the specific enthalpy $h$ and taking the first-order Taylor series expansion of the logarithm around $h \simeq 1$ of Eq.~\eqref{eq:MK_enthalpy} yields
\begin{equation}\label{eq:K_enthalpy}
W - W_{\mathrm{in}} + \frac{K\Gamma}{\Gamma-1}w^{\Gamma-1} + \frac{q}{q-1}K_{\mathrm{m}}(\mathcal{L}w)^{q-1} = 0\,,
\end{equation}
which is the equation for $w$ obtained by~\cite{Komissarov2006a} and it can be solved algebraically.

\begin{table*} 
\begin{center}
\caption{Summary of some relevant quantities of our different models, namely: 
the magnetisation parameter at the centre of the disc $\beta_{\mathrm{m,c}}$, the location of the maximum of the rest-mass density $r_{\mathrm{max}}$, the radial location of the outer boundary of the disc at the equatorial plane $r_{\mathrm{out}}$ (the inner boundary is at $r_{\mathrm{in}}$=1.25 for all discs), the maximum of the rest-mass density of the initial data $\rho_{\mathrm{max,0}}$ (adjusted for a disc initial mass of $M = 0.1 M_{\mathrm{BH}}$), the mass of the disc at the initial time $M_{\mathrm{disc, 0}}$, the maximum of the rest-mass density at the end of our simulation $\rho_{\mathrm{max,F}}$, and the final mass of the disc $M_{\mathrm{disc, F}}$.} 
\begin{tabular}{lcccccccc}
\hline\hline
 Model     &  $\beta_{\rm m,c}$ & $r_{\mathrm{max}}$ & $r_{\mathrm{out}}$ & $\rho_{\mathrm{max,0}}$  &$M_{\mathrm{disc, 0}}$& $\rho_{\mathrm{max,F}}$  &$M_{\mathrm{disc, F}}$\\
\hline
\hline
\texttt{MFD-A} & $10^{3}$  & $1.99$ & $209.0$ & $4.47\times10^{-4}$& $0.1000$ &  $7.78\times10^{-4}$ &$0.0737$ \\
\texttt{MFD-B} & $10^{1}$  & $1.99$ & $209.0$ & $4.47\times10^{-4}$& $0.1000$ &  $4.66\times10^{-4}$ &$0.0744$ \\
\texttt{MFD-C} & $10^{-1}$ & $1.99$ & $209.0$ & $4.47\times10^{-4}$& $0.1000$ &  $3.01\times10^{-5}$ &$0.0084$ \\
\texttt{MFD-D} & $10^{-3}$ & $1.99$ & $209.0$ & $4.47\times10^{-4}$& $0.1000$ &  $4.47\times10^{-9}$ &$0.0000$ \\
\hline
\hline
\texttt{KMh-A} & $10^{3}$  & $1.99$ & $37.40$ & $1.71\times10^{-3}$& $0.1000$ &  $1.51\times10^{-3}$ &$0.0911$ \\
\texttt{KMh-B} & $10^{1}$  & $1.92$ & $36.41$ & $1.71\times10^{-3}$& $0.0643$ &  $1.33\times10^{-3}$ &$0.0592$ \\
\texttt{KMh-C} & $10^{-1}$ & $1.57$ & $28.97$ & $2.99\times10^{-3}$& $0.0234$ &  $7.31\times10^{-4}$ &$0.0078$ \\
\texttt{KMh-D} & $10^{-3}$ & $1.54$ & $28.21$ & $3.41\times10^{-3}$& $0.0237$ &  $5.07\times10^{-4}$ &$0.0054$ \\
\hline
\hline
\texttt{KM$\rho$-A} & $10^{3}$ & $1.99$ & $37.40$ & $1.84\times10^{-3}$& $0.100$ &  $2.12\times10^{-3}$  &$0.0902$ \\
\texttt{KM$\rho$-B} & $10^{1}$ & $1.92$ & $36.41$ & $1.83\times10^{-3}$& $0.071$ &  $1.69\times10^{-3}$ &$0.0656$ \\
\texttt{KM$\rho$-C} & $10^{-1}$ & $1.57$ & $28.97$ & $3.25\times10^{-3}$& $0.025$ &  $8.37\times10^{-4}$ &$0.0083$ \\
\texttt{KM$\rho$-D} & $10^{-3}$ & $1.54$ & $28.21$ & $3.63\times10^{-3}$& $0.025$ &  $5.57\times10^{-4}$ &$0.0058$ \\
\hline
\hline
\end{tabular}
\end{center}
\label{tab:1}
\end{table*}

It is interesting to make a few remarks concerning the validity of the approximation. First of all, we can neglect the magnetic field (i.e.~$K_{\mathrm{m}} \rightarrow 0$) to obtain the non-magnetised fluid approximation. In this case, we can see that the specific enthalpy can be written as
\begin{equation}\label{eq:K_app_enthalpy}
h = 1 + |\Delta W|.
\end{equation}
This result can be considered as the first-order Taylor series approximation of Eq.~\eqref{eq:FD_final_enthalpy}. Then, this shows that, for a non-magnetised flow, $h \simeq 1$ is valid only for small values of $|\Delta W|$. This is not a source of concern as the upper bound\footnote{The upper bound of $|\Delta W|$ is achieved for a Keplerian angular momentum at the radius of the marginally bound orbit, $l = l_{\mathrm{K}}(r_{\mathrm{mb}})$ and $r_{\mathrm{in}} = r_{\mathrm{mb}}$ (this implies $W_{\mathrm{in}} = 0$).} for $|\Delta W|$ goes from $|\Delta W| \simeq 0.0431$ for a Schwarzschild black hole ($a = 0$) to $|\Delta W| = \frac{1}{2} \ln 3 \simeq 0.549$ for a extremal Kerr black hole ($a = 1$)~\citep{Abramowicz78}. Conversely, for a strongly magnetised disc, $p_{\mathrm{m}} \gg p$, it is easy to see that no approximation is done, and this also could be seen as $h \rightarrow 1$ when $K \rightarrow 0$. This shows that the non-relativistic fluid approximation is always valid for strong enough magnetised flows (irrespective of the value of the total potential well $|\Delta W|$).

\begin{figure*}
\includegraphics[width=1.5\columnwidth]{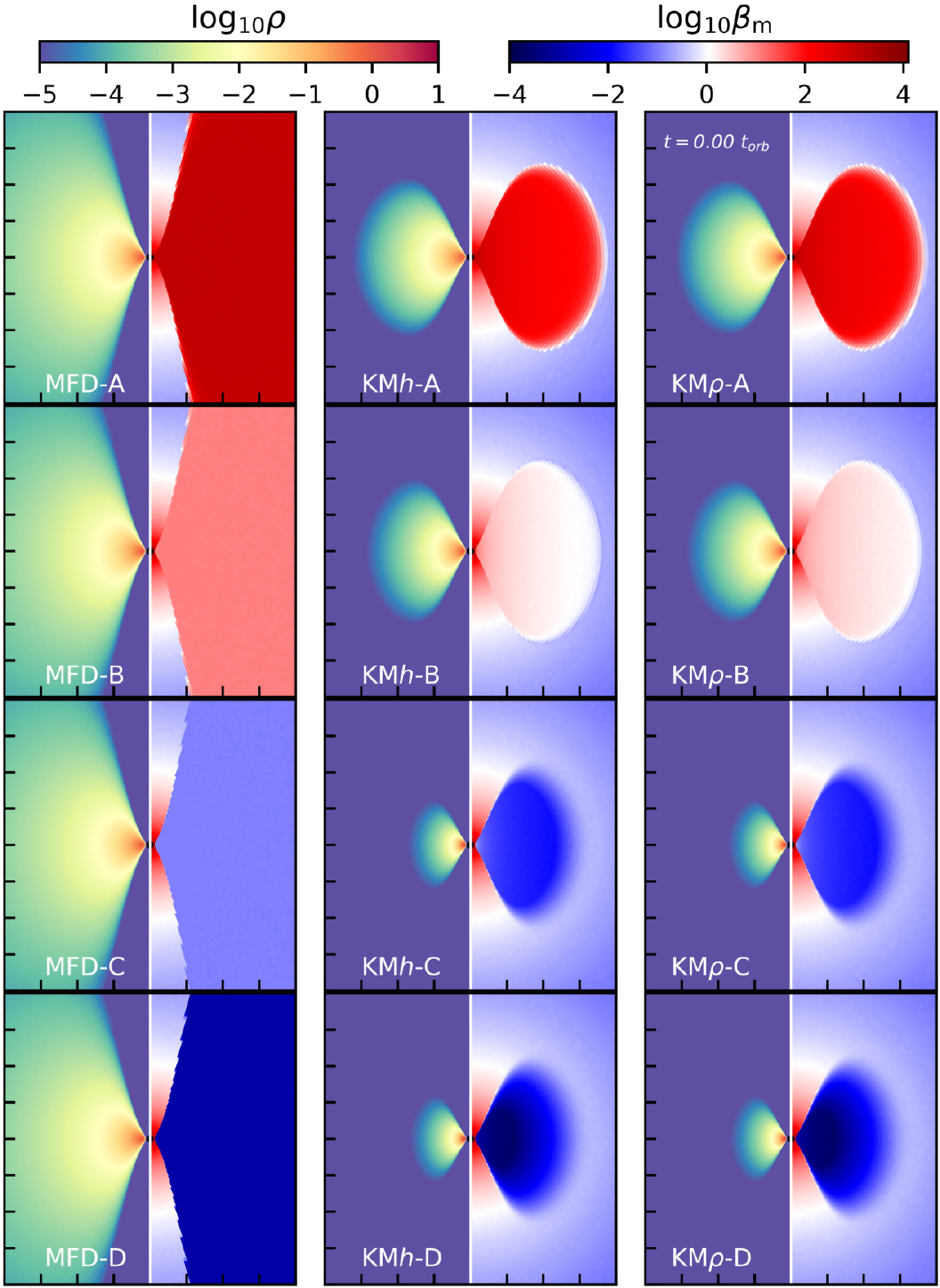}
\caption{ Initial morphology of the rest-mass density (left side of each panel) and magnetisation parameter $\beta_{\mathrm{m}}$ (right side of each panel) for our sample of magnetised tori around a Kerr black hole with spin $a=0.99$ (black circle). From left to right, the columns correspond to models built following the \texttt{MFD}, \texttt{KM}$h$, and \texttt{KM}$\rho$ approaches, respectively. From top to bottom, the rows correspond to models with different values of the magnetisation parameter $\beta_{\mathrm{m,c}}$, namely $10^{3}, ~10^{1},~10^{-1}$, and $~10^{-3}$. The domain plotted on each panel corresponds to $(x,z) \in [-40M, 40M] \times [-40M,40M]$. For models \texttt{KM}$h$ and \texttt{KM}$\rho$ the discs are smaller and the maximum of the density is further inward  the lower the value of $\beta_{\mathrm{m,c}}$. Models \texttt{MFD} do not show such dependence as they are purely hydrodynamical initially.}
\label{fig:Initial2D}
\end{figure*}

\begin{figure}
\begin{center}
\includegraphics[width=1.0\columnwidth]{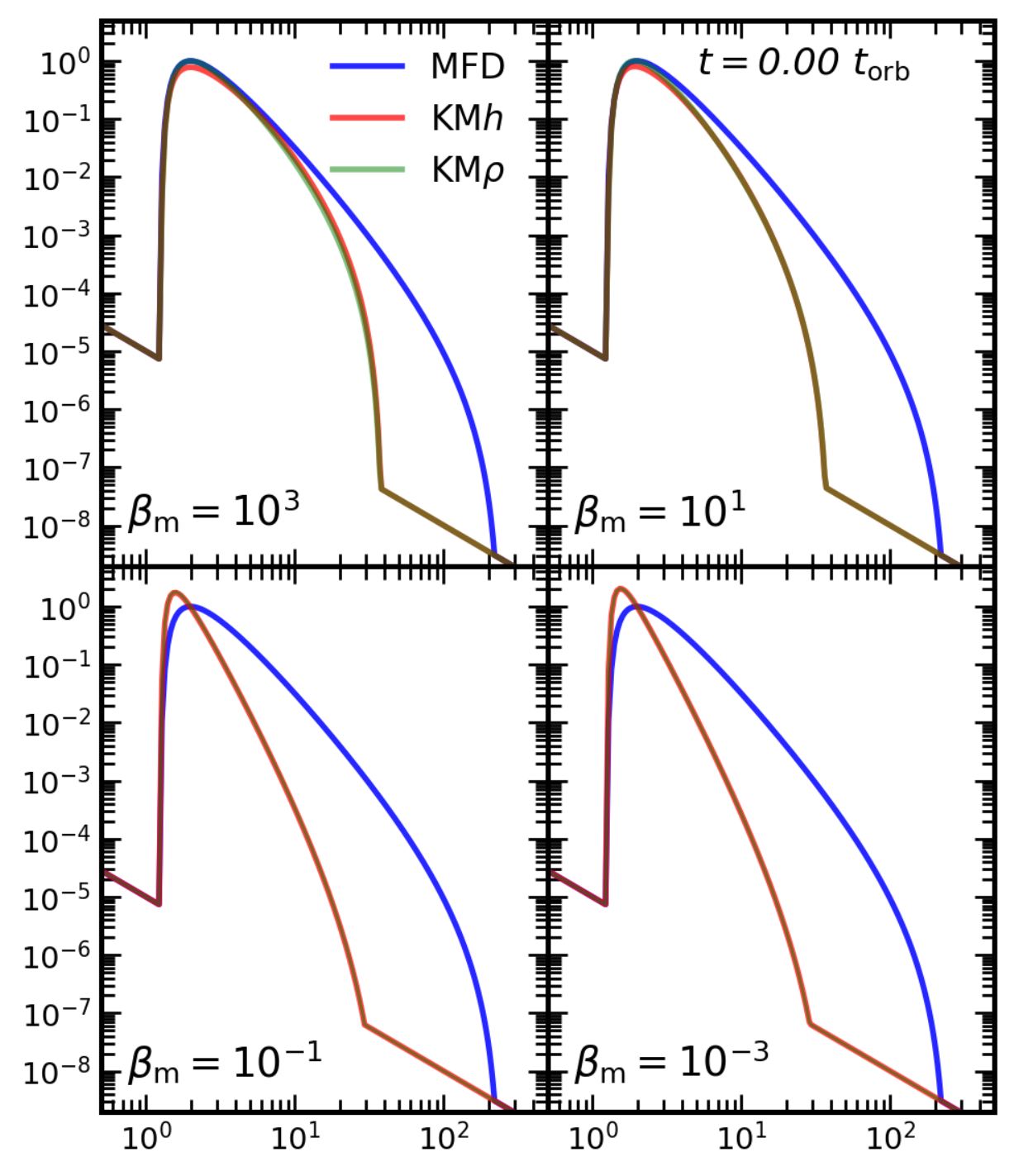}
\caption{ Radial profiles of the rest-mass density at the equatorial plane at the initial time for the four values of the magnetisation parameter considered, indicated in the legends. Blue, red and green lines correspond to approaches  \texttt{MFD}, \texttt{KM}$h$ and \texttt{KM}$\rho$ to build the initial data, respectively. The radial extent of the \texttt{MFD} discs does not depend on $\beta_{\mathrm{m}}$ and is significantly larger than that of the two other approaches that incorporate the magnetic field in a self-consistent way.  \texttt{KM}$h$ and \texttt{KM}$\rho$ discs are hardly distinguishable, becoming practically identical in the most highly magnetised cases (the red and green lines overlap in the bottom panels).}
\label{fig:1DDensity_initial}
\end{center}
\end{figure}

\subsection{Parameters and construction of the discs}

In order to build the discs we have to choose a suitable set of parameters for each one of the three approaches.
For the \texttt{MFD} model (and following~\cite{Font02a}) we fix the specific enthalpy at the inner edge of the disc as $h_{\mathrm{in}} = 1$ and the polytropic constant as $K = 1.5 \times 10^{20} \mathrm{cgs}$. The free parameters for this approach are the adiabatic exponent $\Gamma$, the radial coordinate of the inner edge of the disc $r_{\mathrm{in}}$ (and thus, the total potential at the inner edge of the disc $W_{\mathrm{in}}$), and the specific angular momentum $l$. For the \texttt{KM$\rho$} model, we fix the rest-mass density at the centre $\rho_{\mathrm{c}}$ as $\rho_{\mathrm{c}} = 1$ and we also set the exponent of the magnetic pressure EoS equal to the exponent of the fluid pressure EoS, $q = \Gamma$. The free parameters for this approach are then $\Gamma$, $r_{\mathrm{in}}$, $l$, and the magnetisation parameter at the centre of the disc, $\beta_{\mathrm{m,c}}$. Finally, for the \texttt{KM$h$} model, we proceed as for the \texttt{KM$\rho$} model but fixing the fluid enthalpy at the centre, $w_{\mathrm{c}} = 1$.

For the sake of simplifying the comparison between the three approaches, we fix most of the parameters that characterise the discs and only vary the value of the magnetisation parameter. Therefore, our discs are described by the following set of parameters: the polytropic exponent, which is set to $\Gamma = 4/3$, the constant specific angular momentum, which is set to the value of the Keplerian angular momentum at the marginally bound orbit $l = l_{\mathrm{K}}(r_{\mathrm{mb}})=2.2$, the radial coordinate of the inner edge of the disc, which is chosen to be such that $W_{\mathrm{in}} = 0.1 W_{\mathrm{c}}$ and leads to $r_{\mathrm{in}}=1.25$ (then, the potential gap is set to $\Delta W=0.222$), the black hole spin, set to $a=0.99$, the radius of the cusp,  $r_{\mathrm{cusp}}=1.21$, and the radius of the centre of the disc, $r_{\mathrm{c}}=1.99$. We furthermore introduce a dynamical timescale given by the orbital period measured at the centre of the tori, $t_{\mathrm{orb}}=23.86$. In total we build and evolve 12 equilibrium models, corresponding to the three ways to construct the initial data for magnetised tori, namely \texttt{MFD}, \texttt{KM}$h$, and  \texttt{KM}$\rho$, and four different values of the magnetisation parameter at the centre of the disc, $\beta_{\rm m,c}=10^{3}, 10, 10^{-1}$, and $10^{-3}$, cases $A, ~B,~ C$ and $D$, respectively. Case $A$ corresponds to a weakly magnetised disc (i.e., nearly purely hydrodynamical) and case $D$ is a highly magnetised torus, while the other two are intermediate cases. Numerical values for some relevant quantities characterising the 12 disc models are reported in Table~1. In order to test the dynamics of the tori, we apply a $4 \%$ perturbation on the thermal pressure, namely we use $p = p (1 + 0.04 \chi_r)$, where $\chi_r = (2r_i-1)$ and $r_i$ is a random number. We note that, while the discs do not completely fill their corresponding Roche lobe, the addition of this perturbation is enough to trigger accretion.

\section{Results}
\label{sec:Results}

\subsection{Initial data}

\begin{figure}
\begin{center}
\includegraphics[width=1.0\columnwidth]{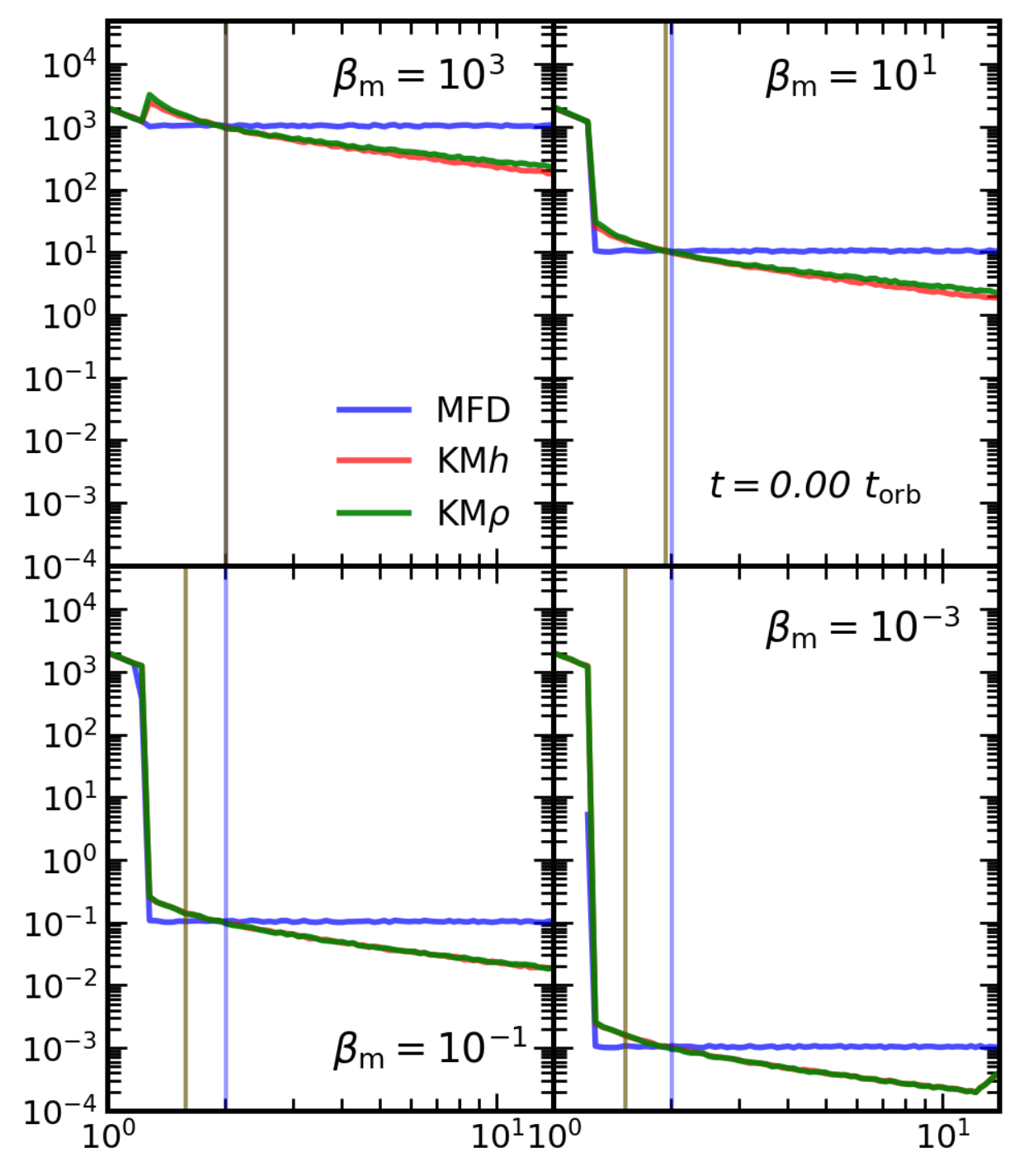}
\caption{ Radial profiles of the disc magnetisation at the equatorial plane at the initial time for the four values of the magnetisation parameter considered, indicated in the legends. Blue, red and green lines correspond to approaches  \texttt{MFD}, \texttt{KM}$h$ and \texttt{KM}$\rho$, respectively. Whereas for \texttt{MFD} discs, $\beta_{\mathrm{m}}$ is constant along the disc, for approaches \texttt{KM}$h$ and \texttt{KM}$\rho$ the magnetisation parameter distribution follows Eq.~\eqref{eq:beta_enth_h} and Eq.~\eqref{eq:beta_enth}, respectively. The vertical lines indicate the location of the maximum of the rest-mass density for each disc using the same color code. Beyond this maximum, \texttt{KM}$h$ and \texttt{KM}$\rho$ discs are significantly more magnetised than \texttt{MFD} discs.}
\label{fig:1DBeta_initial}
\end{center}
\end{figure}

\begin{figure}
\begin{center}
\includegraphics[width=1.0\columnwidth]{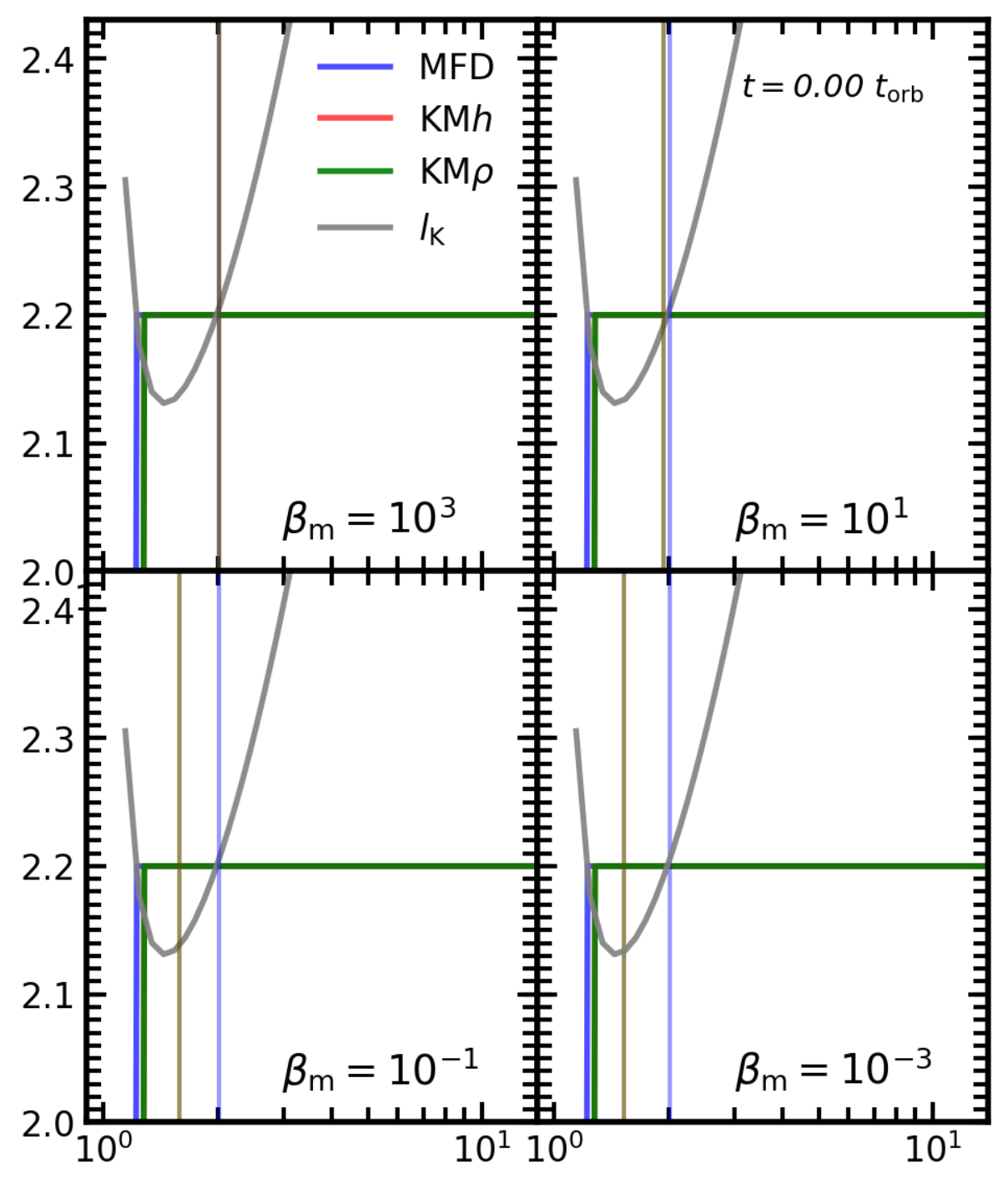}
\caption{ Radial profiles of the specific angular momentum at the equatorial plane at the initial time for the four values of the magnetisation parameter considered, indicated in the legends. Blue, red and green lines correspond to approaches  \texttt{MFD}, \texttt{KM}$h$ and \texttt{KM}$\rho$, respectively, and the vertical lines show the location of the maximum of the rest-mass density for each disc using the same color code. The specific angular momentum is initially constant and equal for all models, by construction. With respect to the Keplerian angular momentum profile, depicted by the grey line, all discs are composed of an inner super-Keplerian region, $[r_{\rm in}, r_{\rm c})$, and an outer sub-Keplerian region, $(r_{\rm c}, r_{\rm out}]$.}
\label{fig:1DAngMom_initial}
\end{center}
\end{figure}

We start by discussing the initial data of the 12 disc models we are going to evolve. These models are depicted in Figs.~\ref{fig:Initial2D} to \ref{fig:1DAngMom_initial}, which display the 2D morphology of the discs (Fig.~\ref{fig:Initial2D}) and the radial profiles of selected quantities on the equatorial plane (Figs.~\ref{fig:1DDensity_initial} to \ref{fig:1DAngMom_initial}).

Fig.~\ref{fig:Initial2D} shows the logarithm of the rest-mass density and the logarithm of the magnetisation parameter for our sample of 12 initial models. Each row of this figure corresponds to a different value of the magnetisation parameter at the centre of the disc and each column indicates one of the three approaches we use to construct the magnetised discs. We note that, despite the atmosphere has no magnetic field, in order to plot the magnetisation $\beta_{\rm m}=p/p_{\rm m}$ we need to select a non-zero value of the magnetic pressure (namely, $p_{\rm m}=10^{-10}$). This explains the spherical distribution of the magnetisation parameter visible outside the discs in figure \ref{fig:Initial2D} (and in Fig.~\ref{fig:rhobeta2Dt} below).

For models \texttt{MFD-A} to \texttt{MFD-D} (left panel of the first column), the rest-mass density and all thermodynamical quantities are identical due to the fact that the initial data for these models are purely hydrodynamical at first (\ie at $t = 0$, the fluid pressure and the magnetic pressure distributions do not \textit{see} each other and the interaction between them is introduced from the first timestep onward). By contrast, the morphology of models \texttt{KM}$h$ and \texttt{KM}$\rho$ (middle and right columns) changes for varying values of $\beta_{\mathrm{m, c}}$. In particular, the size of the disc is smaller for lower values of the magnetisation parameter (i.e.~a stronger magnetic field) at the centre $\beta_{\mathrm{m,c}}$, and the location of the maximum of the density moves towards the inner edge of the disc. This can be better observed in Fig.~\ref{fig:1DDensity_initial}, where we plot the logarithm of the rest-mass density versus the logarithm of the radial coordinate at the equatorial plane. The radial location of the outer boundary of the discs along the equatorial plane is reported in Table 1. We note that the maximum of the rest-mass density for models \texttt{KM}$h$\texttt{-A} and \texttt{KM}$h$\texttt{-B} is less than one. The reason is because, in this approach, we set $w_{\mathrm{c}} = 1$, and then $\rho_{\mathrm{c}} = w_{\mathrm{c}}/h_{\mathrm{c}}$. It follows from Eq.~\eqref{eq:K_app_enthalpy} that $\rho_{\mathrm{c}} = 1/(1 - |W_{\mathrm{c}}-W_{\mathrm{in}}|)<1$. The interested reader is addressed to~\cite{Gimeno-Soler:2017} and~\cite{Gimeno-Soler:2019} for an extensive discussion on the morphology of magnetised discs for different degrees of magnetisation.

\begin{figure*}
\includegraphics[width=1.5\columnwidth]{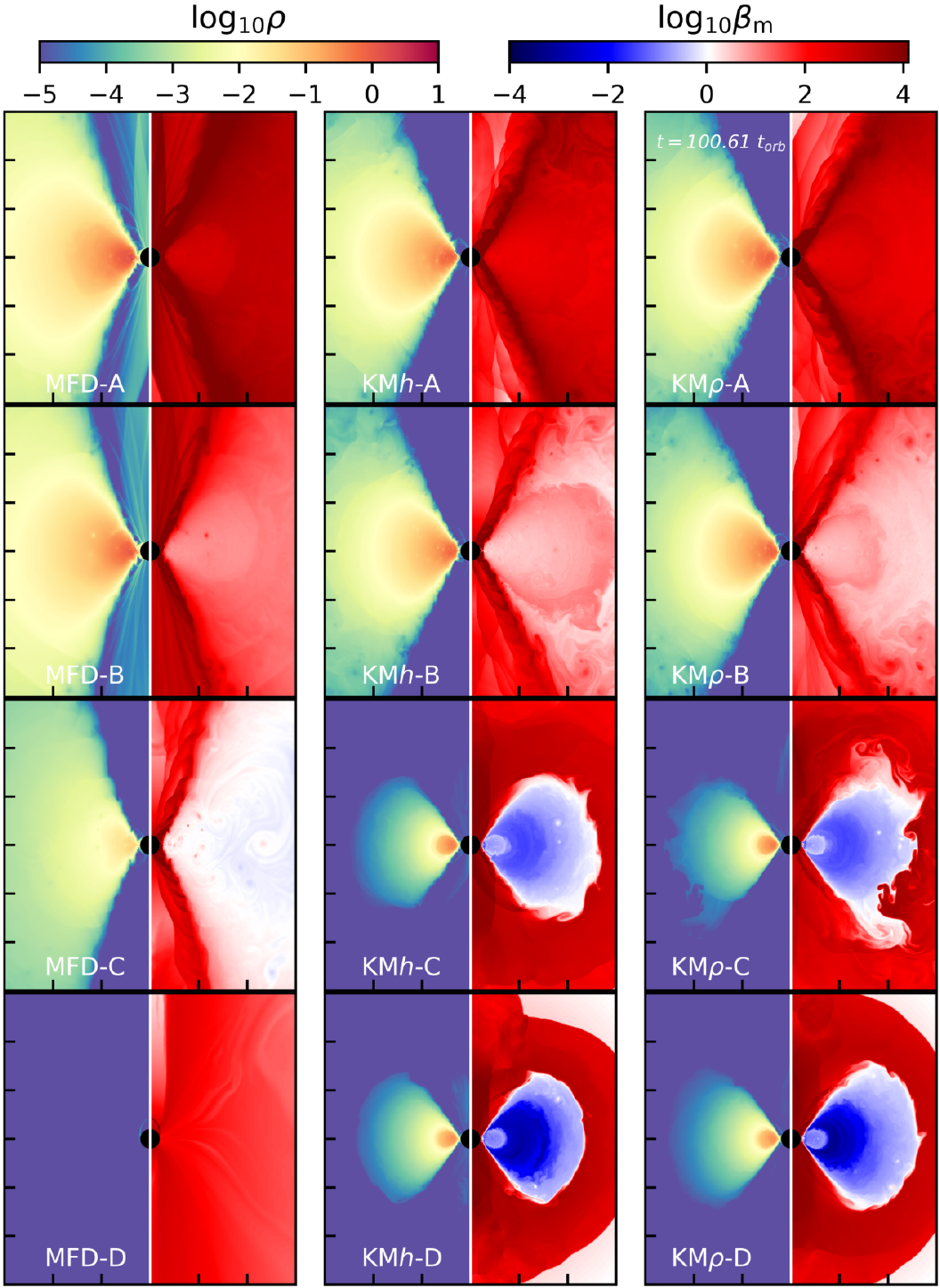}
\caption{ Final morphology (at $t \sim 100 t_{\rm orb}$) of the rest-mass density (left side of each panel) and magnetisation parameter $\beta_{\mathrm{m}}$ (right side of each panel) for our sample of magnetised tori around a Kerr black hole with spin $a=0.99$. From left to right, the columns correspond to models built following the \texttt{MFD}, \texttt{KM}$h$, and \texttt{KM}$\rho$ approaches, respectively. From top to bottom, the rows correspond to models with different values of the magnetisation parameter $\beta_{\mathrm{m,c}}$, namely $10^{3}, ~10^{1},~10^{-1}$, and $~10^{-3}$. The domain plotted on each panel corresponds to $(x,z) \in [-15M, 15M] \times [-15M,15M]$. For low magnetisation values ($\beta_{\mathrm{m,c}} = 10^3$ and 10), the final rest-mass density distribution of the discs is similar for the three approaches but the \texttt{MFD} discs become less magnetised than the other two. However, for high magnetisation values the \texttt{MFD} discs are significantly perturbed to even become completely disrupted for $\beta_{\mathrm{m,c}} = 10^{-3}$.  \texttt{KM}$h$, and \texttt{KM}$\rho$ discs remain stable throughout although they become significantly smaller compared to their original size, due to accretion.}
\label{fig:rhobeta2Dt}
\end{figure*} 

\begin{figure}
\begin{center}
\includegraphics[width=1.0\columnwidth]{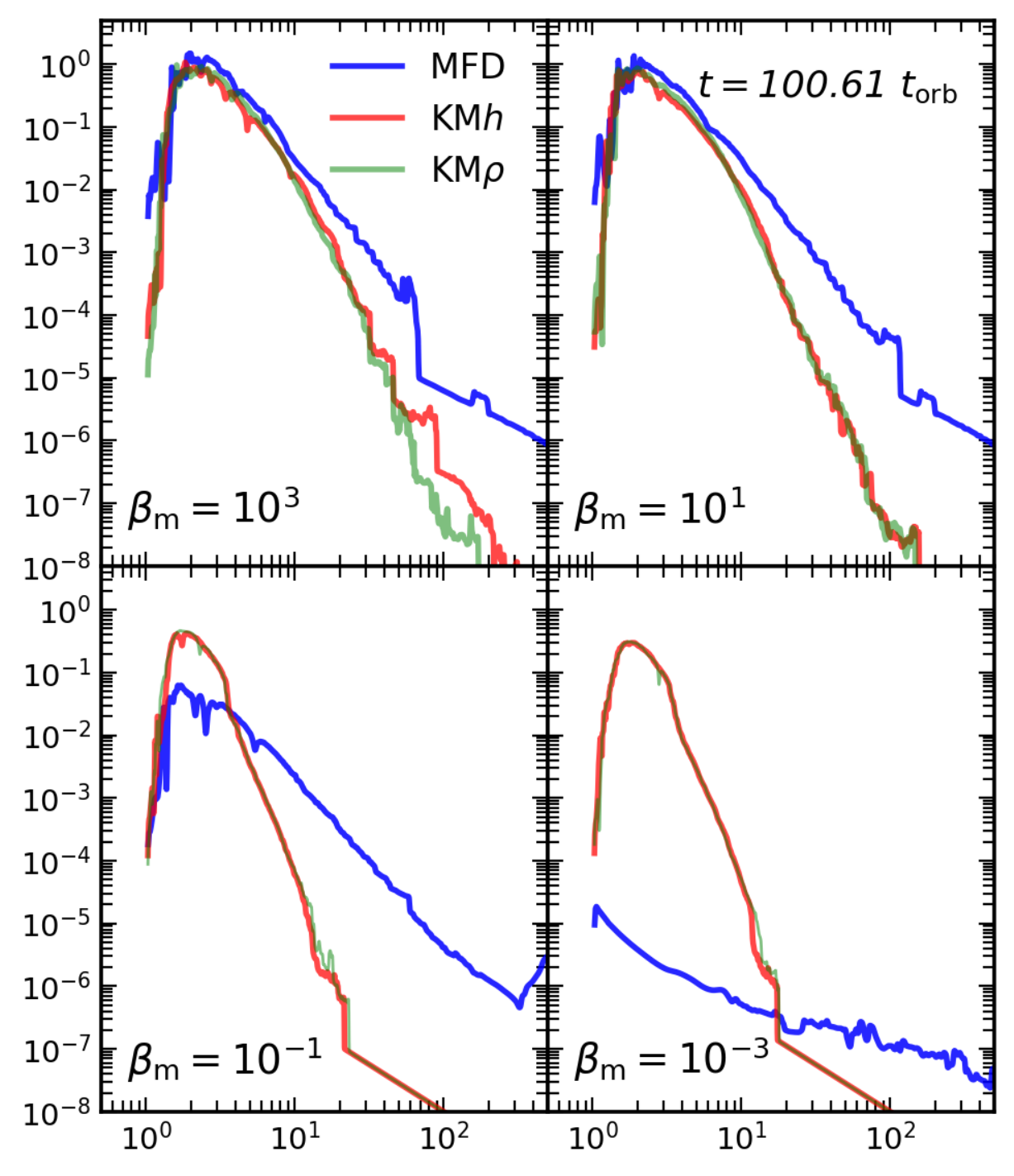}
\caption{Radial profiles of the rest-mass density at the equatorial plane at the end of the evolution ($100.61 t_{\mathrm{orb}}$)
for the four values of the magnetisation parameter considered, indicated in the legends. Blue, red and green lines correspond to approaches  \texttt{MFD}, \texttt{KM}$h$ and \texttt{KM}$\rho$, respectively. Comparing with the initial profiles (Fig.~\ref{fig:1DDensity_initial}) the general shape of the discs is preserved during the evolution, except for model \texttt{MFD-D} (blue line at the bottom-right panels) where the disc is destroyed. The agreement between approaches \texttt{KM}$h$ and \texttt{KM}$\rho$ is also maintained.}
\label{fig:1DDensity_final}
\end{center}
\end{figure}

\begin{figure}
\begin{center}
\includegraphics[width=1.0\columnwidth]{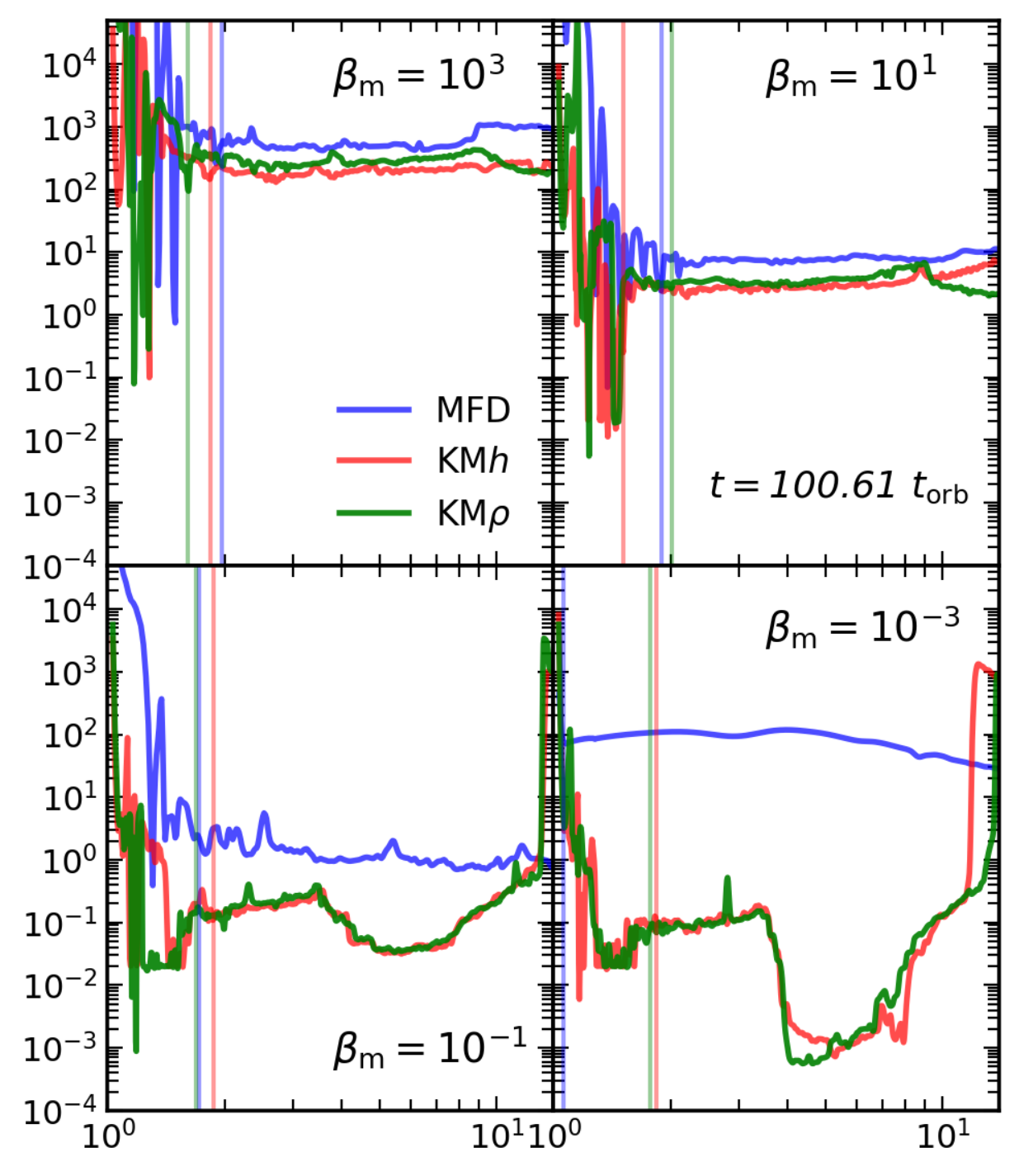}
\caption{ Radial profiles of the disc magnetisation at the equatorial plane at the end of the evolution ($100.61 t_{\mathrm{orb}}$)
for the four values of the magnetisation parameter considered, indicated in the legends. Blue, red and green lines correspond to approaches  \texttt{MFD}, \texttt{KM}$h$ and \texttt{KM}$\rho$, respectively, and the vertical lines show the location of the maximum of the rest-mass density for each disc using the same color code. For weakly magnetised disks (top panels) the magnetisation is roughly constant along the disc and its value only increases slightly with respect to its initial value (cf.~Fig.~\ref{fig:1DBeta_initial}. For strongly magnetised disks (bottom panels), and for approaches \texttt{KM}$h$ and \texttt{KM}$\rho$, we observe the development of a highly magnetised envelope surrounding the high-density central region of the disc.}
\label{fig:1DBeta_final}
\end{center}
\end{figure}

 In Fig.~\ref{fig:1DBeta_initial} we show the 1D initial profiles of the magnetisation at the equatorial plane for each procedure and also the location of the maximum of the rest-mass density, indicated by the vertical lines. For the purely hydrodynamical solutions \texttt{MFD} the location of this maximum is at the centre of the disc (vertical blue line). As it can be seen, the behaviour of $\beta_{\rm m}$ is different for the models \texttt{MFD} on the one hand and for the models \texttt{KM}$h$ and \texttt{KM}$\rho$ on the other hand. This is expected, as the method to build the magnetic field is different. In particular, as we mentioned before, for the \texttt{MFD} approach $\beta_{\rm m}$ is constant throughout the disc, and for the \texttt{KM}$h$ and \texttt{KM}$\rho$ cases $\beta_{\rm m}$ decreases with increasing radial coordinate. This fact can be easily explained when $\beta_{\rm m}$ is written as
\begin{equation}
\label{eq:beta_enth_h}
\beta_{\rm m} = \frac{K}{K_{\rm m} \mathcal{L}^{\Gamma-1}}\,,
\end{equation}
for the \texttt{KM}$h$ models, and as
\begin{equation}
\label{eq:beta_enth}
\beta_{\rm m} = \frac{K}{K_{\rm m} h^{\Gamma}\mathcal{L}^{\Gamma-1}}\,,
\end{equation}
for the \texttt{KM}$\rho$ models. The presence of the specific enthalpy $h$ in equation~\eqref{eq:beta_enth} also explains the differences observed between the cases \texttt{A} and \texttt{B} for models \texttt{KM}$h$ and \texttt{KM}$\rho$. Additionally, in Fig.~\ref{fig:1DAngMom_initial} we show the initial radial profiles at the equatorial plane of the specific angular momentum, the Keplerian angular momentum and the location of the maximum of the rest-mass density (indicated with blue, red and green vertical lines for models \texttt{MFD}, \texttt{KM}$h$ and \texttt{KM}$\rho$ respectively). By construction, the specific angular momentum is initially constant and we can observe that, with respect to the Keplerian angular momentum, the disc is divided into two regions: A first super-Keplerian region, spanning the interval $[r_{\rm in}, r_{\rm c})$ and a second sub-Keplerian region at $(r_{\rm c}, r_{\rm out}]$.

\subsection{Late time morphology}

\begin{figure}
\begin{center}
\includegraphics[width=1.0\columnwidth]{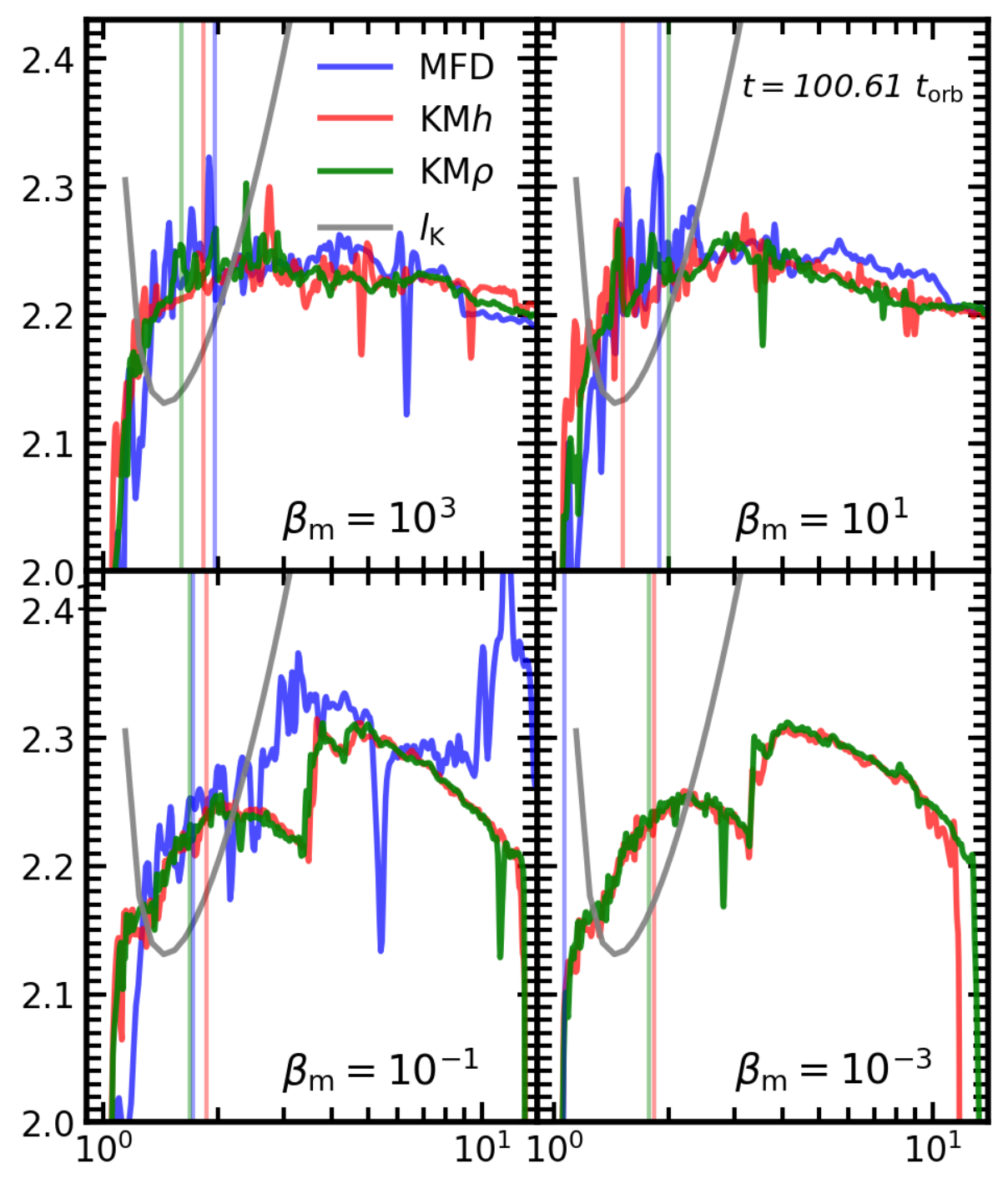}
\caption{ Radial profiles of the specific angular momentum at the equatorial plane at the end of the evolution ($100.61 t_{\mathrm{orb}}$) for the four values of the magnetisation parameter considered, indicated in the legends. Blue, red and green lines correspond to approaches  \texttt{MFD}, \texttt{KM}$h$ and \texttt{KM}$\rho$, respectively, and the vertical lines show the location of the maximum of the rest-mass density for each disc using the same color code. The Keplerian angular momentum is depicted with a grey line. The comparison with the initial profile (cf.~Fig.~\ref{fig:1DAngMom_initial}) shows that the angular momentum drops in the inner regions of the discs for all models and increases slightly above $2.2$ in the rest. The increase is larger for strongly magnetised discs (bottom panels) where an external envelope with a higher value of the angular momentum forms, coinciding with the highly magnetised region observed in Fig.~\ref{fig:1DBeta_final}.}
\label{fig:1DAngMom_final}
\end{center}
\end{figure}

\begin{figure}
\begin{center}
\includegraphics[width=0.9\columnwidth]{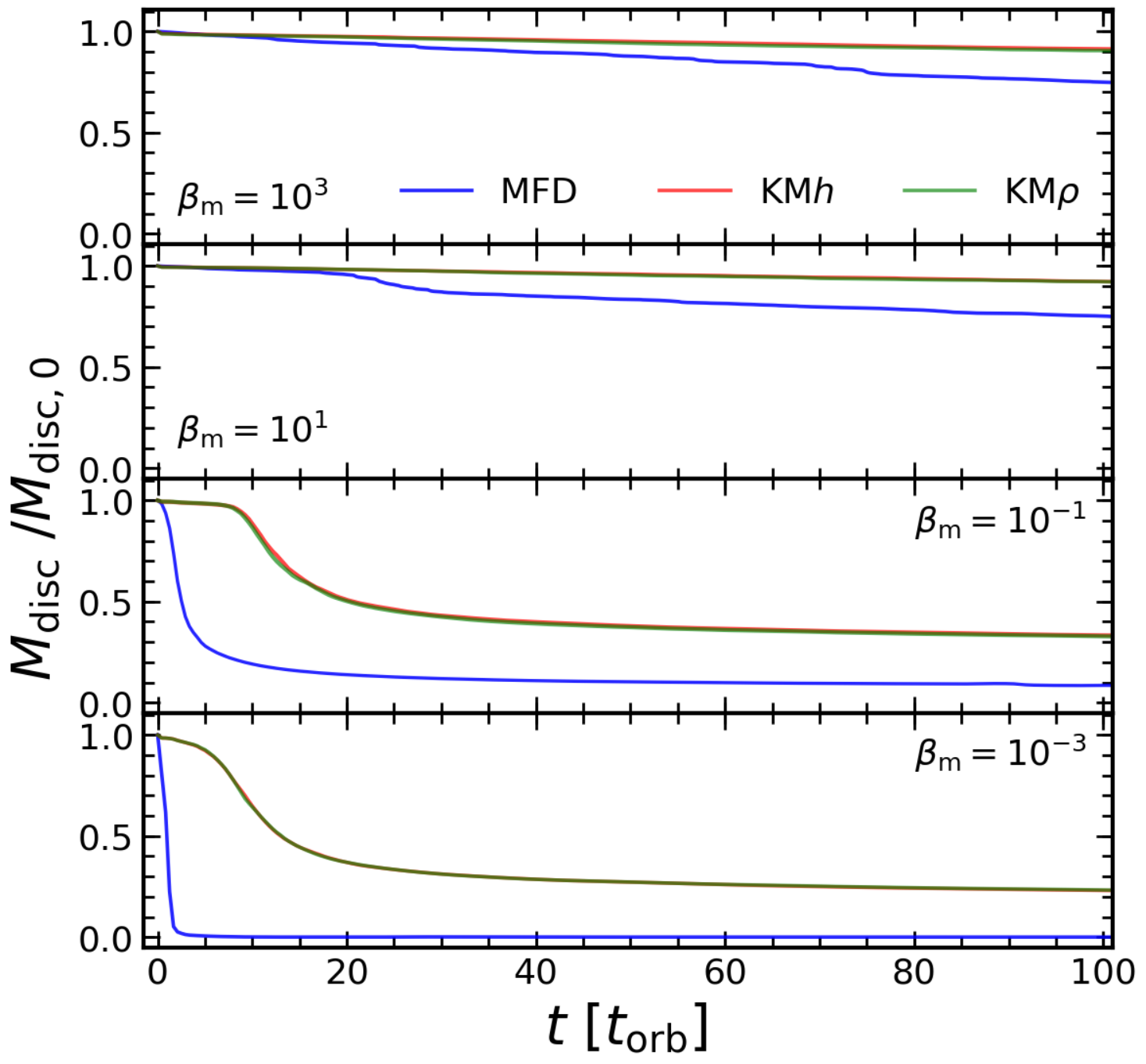}
\caption{ Evolution of the mass of the disc in units of its initial value. From top to bottom, the panels correspond to $\beta_{\mathrm{m}}=10^3$, $10$, $10^{-1}$, and $10^{-3}$. Models \texttt{MFD}, \texttt{KM}$h$ and \texttt{KM}$\rho$ are shown in blue, red and green lines, respectively, in each panel. The initial perturbation triggers the accretion of mass on to the black hole for all models. The effect is more pronounced and rapid as the magnetisation is increased, especially for the \texttt{MFD} discs, where mass is also expelled, for which the final mass drops to a $\sim 10\%$ value of the initial mass for $\beta_{\mathrm{m}}=10^{-1}$ and to negligible values for $\beta_{\mathrm{m}}=10^{-3}$.}
\label{fig:Mass}
\end{center}
\end{figure}

\begin{figure}
\begin{center}
\includegraphics[width=0.9\columnwidth]{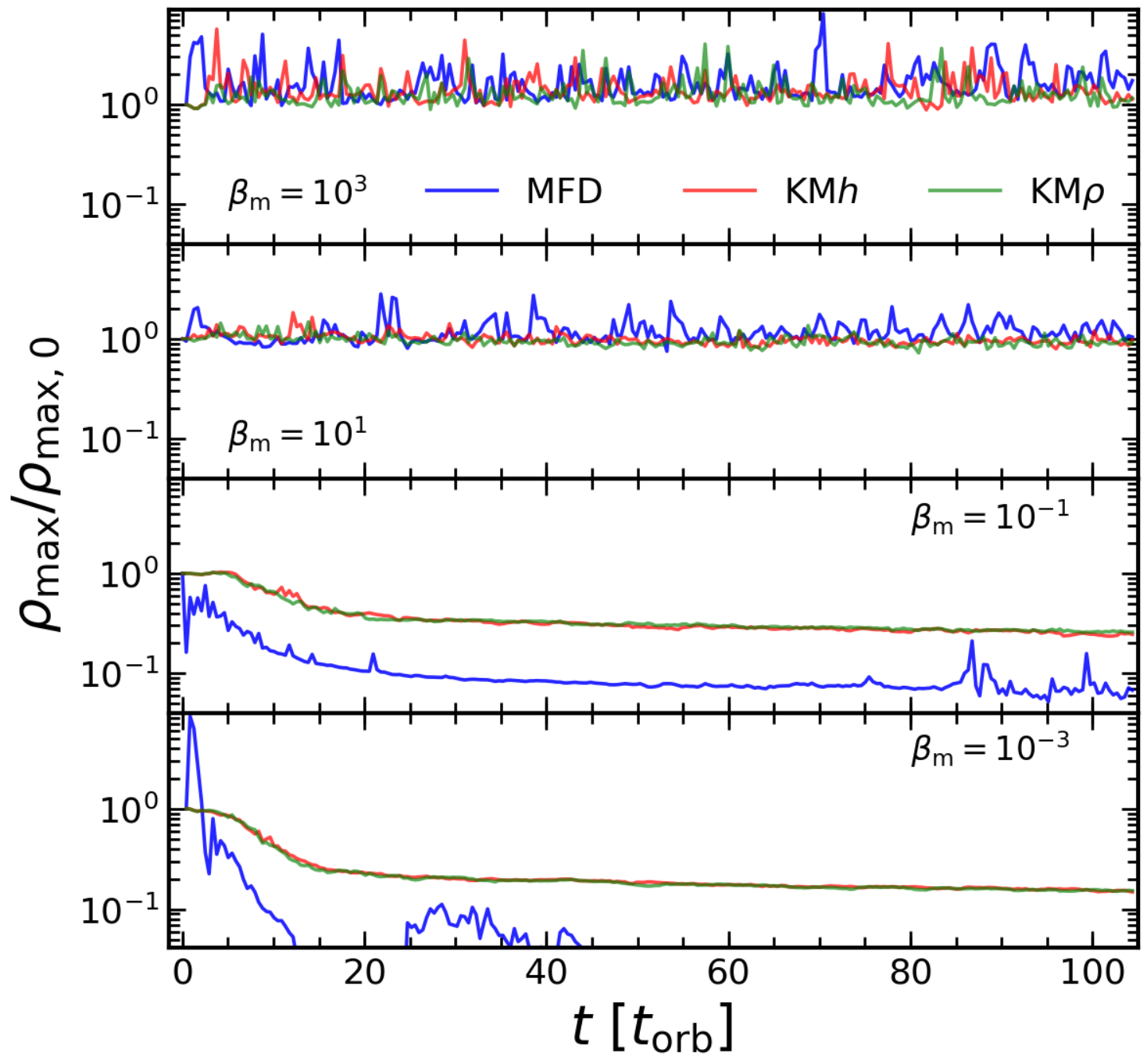}
\caption{ Evolution of the maximum of the rest-mass density at the equatorial plane normalized by its maximum at the initial time. From top to bottom, the panels correspond to $\beta_{\mathrm{m}}=10^3$, $10$, $10^{-1}$, and $10^{-3}$. Models \texttt{MFD}, \texttt{KM}$h$ and \texttt{KM}$\rho$ are shown in blue, red and green lines, respectively, in each panel. The maximum rest-mass density stays close to its initial value for the less magnetised models but drops as the magnetisation is increased, especially for the \texttt{MFD} discs.}%
\label{fig:MassDensity}
\end{center}
\end{figure}

We evolve the initial data for about 100 orbital periods in order to find out whether noticeable long-term differences appear in the discs, both with respect to the initial data and among them, due to the way the magnetic field is set up in the three approaches. The results of the simulations are depicted in Figs.~\ref{fig:rhobeta2Dt} to~\ref{fig:1DAngMom_final}. In addition, Figs.~\ref{fig:Mass} and~\ref{fig:MassDensity} show the time evolution of the fraction of the initial total mass that remains in the disc and the time evolution of the rest-mass density at the centre of the disc normalised by its value at the initial time, respectively.

The late time 2D morphology of the discs is shown in Fig.~\ref{fig:rhobeta2Dt}. As in Fig.~\ref{fig:Initial2D}, the columns correspond to the three different models (namely \texttt{MFD}, \texttt{KM}$h$ and \texttt{KM}$\rho$), the rows correspond to the four values of the magnetisation parameter that we have considered (namely $10^3$, $10$, $10^{-1}$, $10^{-3}$). Likewise, the left half of each panel of Fig.~\ref{fig:rhobeta2Dt} displays the logarithm of the rest-mass density whereas the right half displays the magnetisation parameter in logarithmic scale.

The perturbation of the initial data triggers accretion of the material of the discs on to the black hole. Figure \ref{fig:rhobeta2Dt} shows that for the lowest magnetisation we have considered ($\beta_{\mathrm{m,c}} = 10^3$), the rest-mass density distribution of the discs after the evolution is very similar for the three approaches. In particular, the only perceptible difference is that the disc built using the \texttt{MFD} approach is slightly bigger. Regarding the evolution of the magnetisation, we can see that after 100 orbital periods the discs have undergone a redistribution of their magnetic field, with the appearance of a slightly more magnetised toroidal region (with respect to the initial data values) which coincides with the most dense region of the disc. We note that the disc built with the \texttt{MFD} approach is the less magnetised of the three approaches.

This trend continues when we observe the second row of Fig.~\ref{fig:rhobeta2Dt}  (which corresponds to $\beta_{\mathrm{m,c}} = 10$). In this case, the differences between the \texttt{MFD} and the \texttt{KM} approaches are more apparent: the discs are smaller and more magnetised in both \texttt{KM} cases when compared to the \texttt{MFD} disc. Nevertheless, the morphology of the three discs is still quite similar.

So far, we have compared low magnetised models, so it is expected that the discrepancy between the different approaches should be small. In particular, the addition of the magnetic field for the low-magnetised \texttt{MFD} models introduces only a small perturbation of the fluid pressure (the magnetic pressure). However, when the magnetisation parameter goes below $\beta_{\mathrm{m}} = 1$, the magnetic pressure becomes larger than the fluid pressure and it is not longer possible for it to be considered as a small perturbation.

In the third row of Fig.~\ref{fig:rhobeta2Dt} we consider discs with a magnetisation parameter of $\beta_{\mathrm{m,c}} = 10^{-1}$, Here, we start to see the limits of the \texttt{MFD} approach. As we can observe, the outcome of the evolution is now very different for the \texttt{MFD} and the \texttt{KM}$h$ and \texttt{KM}$\rho$ approaches. In the first case, the disc is significantly bigger, the maximum rest-mass density is significantly lower and it is less magnetised than its \texttt{KM} counterparts. This is caused by the introduction of the magnetic field as a perturbation of the pressure; in this case, the (overall) perturbation is large enough to further intensify the accretion of a significant part of the disc (hence the drop of the maximum of the rest-mass density). Nevertheless, for this value of the magnetisation parameter the shape of the disc is not yet drastically altered.

Finally, in the last row of Fig.~\ref{fig:rhobeta2Dt}, we show the outcome of the evolution of the highest magnetised discs we have considered in this work. In this case, after 100 orbital periods, the disc built using approach \texttt{MFD} has entirely disappeared. The reason is because, as discussed before, the `ad hoc' magnetic field is introduced as a perturbation of the pressure, but compared to the previous case it is now $100$ times bigger (as $\beta_{\mathrm{m}} = 10^{-3}$). Therefore, the magnetisation is sufficiently large to disrupt the disc in such a way that by the end of the simulation the disc material has been either accreted by the black hole or expelled away, leaving behind a low-magnetised remnant hardly distinguishable from the atmosphere. 

It is worth now to describe the 2D morphology of the discs at $t \sim 100 t_{\mathrm{orb}}$ for both \texttt{KM} models and magnetisations $\beta_{\mathrm{m,c}} = 10^{-1}$ and $\beta_{\mathrm{m}} = 10^{-3}$. For these two cases, it is apparent that the final disc is significantly smaller when compared to its intial state. Also, it can be seen that the value of the maximum rest-mass density is also smaller. This is due to the initial perturbation we applied in the pressure. As these highly-magnetised discs have the location of the maximum of the rest-mass density $r_{\mathrm{max}}$ closer to the inner edge of the disc (and hence, closer to the black hole), a perturbation can trigger the accretion of a greater amount of matter in an easier way. The magnetisation distribution of these discs is also different. The central, highest density region ends up becoming less magnetised than at the start of the simulation, whereas the external, less dense layers of the disc are endowed with a stronger magnetic field than at the initial time. Besides this, we can also see that the two \texttt{KM} approaches yield almost the same outcome after the evolution. This is expected, as they coincide when $\beta_{\mathrm{m}} \to 0$.

In Figs.~\ref{fig:1DDensity_final}, \ref{fig:1DBeta_final} and \ref{fig:1DAngMom_final} we show radial slices along the equatorial plane of the rest-mass density, the magnetisation parameter, and the specific angular momentum, respectively. The rest-mass density plots show that for low magnetisation (high values of $\beta_{\mathrm{m}}$) the radial profiles at the end of the evolution closely resemble those at $t=0$, irrespective of the prescription employed to account for the magnetic field. In particular, the peak values and location of the density remain roughly the same of the initial values. (The differences in the location of the density maxima can be appreciated best by comparing the radial position of the vertical lines in Fig.~\ref{fig:1DBeta_final} with their counterparts in Fig.~\ref{fig:1DBeta_initial}. Note that the radial scale is logarithmic.) The most important difference is the formation of a more extended, low-density, envelope for large radii for the three models. However, for the two most highly magnetized models, the late time radial profiles of the density show important differences with the initial profiles. As we also observed in Fig.~\ref{fig:rhobeta2Dt}, while approaches \texttt{KM}$h$ and \texttt{KM}$\rho$ still show disc-like profiles (albeit smaller and the peak density has decreased to $\sim 3 \times 10^{-1}$ for models \texttt{C} and to $\sim 2 \times 10^{-1}$ for models \texttt{D} by $t\sim 100 \,t_{\rm orb}$), the \texttt{MFD-C} model rest-mass profile is more similar to the ones found in less magnetised models but it has suffered a heavy mass loss, with a maximum density at $t\sim 100 \,t_{\rm orb}$ of $\rho_{\mathrm{max}} \sim 7 \times 10^{-2}$, and the \texttt{MFD-D} model has completely vanished, leaving behind only a very low density remnant far from the central black hole. It is also worth noting that, for the two most highly magnetized models, the location of the maximum of the rest-mass density $r_{\mathrm{max}}$ has barely drifted away from the black hole.

The inspection of Fig.~\ref{fig:1DBeta_final} and the comparison with Fig.~\ref{fig:1DBeta_initial} reveals that 
the magnetisation parameter along the disc decreases for weakly magnetised models (\texttt{A} and \texttt{B}) but grows for the stronger magnetised cases (\texttt{C} and \texttt{D}).  This suggests a value of $\beta_{\rm m,c} \sim 1$ for which the magnetisation of the disc is constant during the evolution. The mechanism responsible for the redistribution of the magnetisation is, most likely, the radial compression and expansion that the discs suffer during the evolution. Radial compression of the magnetic field lines lead, in turn, to the local amplification of the magnetic field. On the one hand, for models \texttt{A} and \texttt{B} (and \texttt{MFD-C}), Fig.~\ref{fig:1DDensity_final} shows that, even though the morphology of the discs does not change significantly, the central parts become a little more compact by the end of the evolution. The infall of matter into those central regions produces the corresponding local amplification of the magnetic field (as observed in Fig.~\ref{fig:rhobeta2Dt}). On the other hand, for the two most magnetised models \texttt{C} and \texttt{D}, the dynamics leads to the appearance of two distinct regions in the disc (best visible for $\beta_{\rm m}=10^{-3}$; see also Fig.~\ref{fig:rhobeta2Dt}): a central less magnetised region surrounded by a region where the magnetic field has been slightly amplified.

Radial profiles of the final angular momentum along the equatorial plane are depicted in Fig.~\ref{fig:1DAngMom_final}. This figure reveals a redistribution of the specific angular momentum along the disc. For cases \texttt{A} and \texttt{B} (and case \texttt{MFD-C}), and irrespective of the approach used to incorporate the magnetic field, the specific angular momentum increases slightly in the central region of the disc but overall stays roughly constant. However, for the \texttt{KM} models \texttt{C} and \texttt{D} we end up with a different configuration: a lower $l$ region near the inner edge of the disc as in the previous cases, a second high-density region with $l > 2.2$ containing a local maximum of the specific angular momentum, and a third low-density region, also with $l > 2.2$, which contains the absolute maximum of the specific angular momentum. In any event, the specific angular momentum does not change much during the evolution (at most $\pm \sim5\%$). Moreover, it can be seen that the initial structure (an inner super-Keplerian region and an outer sub-Keplerian region) is preserved during the evolution. 

The change in the location of the maximum of the rest-mass density for models \texttt{KM}$h$\texttt{-C/D} and \texttt{KM}$\rho$\texttt{-C/D} is worth a further comment. The fact that a non-constant angular momentum region develops during the evolution of these models and that the inner region of the discs lose part of their magnetisation are the reason of said drift in $r_{\mathrm{max}}$. That can be seen in the central panel of Fig.~(6) shown in~\cite{Gimeno-Soler:2017}, where the authors plot $r_{\mathrm{max}}$ vs.~$\log_{10}\beta_{\mathrm{m,c}}$ for a Kerr black hole with spin parameter $a = 0.99$ and for three different non-constant angular momentum distributions (the black line in that figure represents our \texttt{KM}$h$ models). As it can be seen in~\cite{Gimeno-Soler:2017}, the drop of the magnetisation would not be enough to achieve the values of $r_{\mathrm{max}}$ we observe here; we would need a change in the specific angular-momentum distribution as well.

In Fig.~\ref{fig:Mass} and Fig.~\ref{fig:MassDensity} we show the time evolution of the mass of the discs and of the maximum of the rest-mass density (normalised by the initial values). The disc mass is computed using
\begin{equation}\label{eq:mass}
m = \int \sqrt{\gamma} W \rho \,d^{3}x\,.
\end{equation}
The values of these two quantities, at the initial and final times, are reported in Table~1.
We find that, for the three approaches and the lowest magnetised models (\ie~\texttt{A} and \texttt{B}), the maximum of the rest-mass density is oscillating, but remains close to its initial value. With respect to the fraction of the mass in the disc, we see that the initial perturbation triggers the accretion of a very small fraction of the total mass for the \texttt{KM} models ($> 90\%$ of the mass survives after an evolution of $t\sim 100 \,t_{\rm orb}$). A bit less (about a $75\%$ of the total mass) survives for models \texttt{MFD-A} and \texttt{MFD-B},which we attribute to the presence of an additional source of perturbation due to the inconsistent incorporation of the magnetic field on top of purely hydrodynamic initial data.

By increasing the magnetisation, these trends become more acute. For models \texttt{KM}$h$\texttt{-C} and \texttt{KM}$\rho$\texttt{-C} the maximum of the rest-mass density drops to a $\sim15\%$ fraction of $\rho_{\mathrm{max,0}}$ and the total mass of the disc drops to $\sim33\%$ of the initial mass. The change is more dramatic for the model \texttt{MFD-C}, where the final density is a $\sim 7\%$ fraction of $\rho_{\mathrm{max,0}}$ and the final mass is a $\sim 10\%$ fraction of the initial mass. Again, this is due to the perturbation introduced by the magnetic field being too large. Lastly, for the highest magnetised case, the discs in models \texttt{KM}$h$\texttt{-D} and \texttt{KM}$\rho$\texttt{-D} lose even more matter: the maximum of the rest-mass density is a $\sim 15\%$ of its initial value and the final mass is around $\sim 23\%$ of the initial mass. By recalling the results from our resolution tests in Fig.~\ref{fig:resolution} we note that, at our fiducial resolution, we are overestimating the mass loss for the highly-magnetised \texttt{KM} models by about 10\%. For the \texttt{MFD-D} model, Figs.~\ref{fig:Mass} and \ref{fig:MassDensity} reveal that the disc is rapidly destroyed at the beginning of the evolution, as the total mass drops to negligible values during the first orbital periods. The maximum of the rest-mass density also vanishes but at a different rate, as the code keeps track of the matter that is being expelled away.

\section{Discussion}
\label{sec:Sum}

In this paper we have built equilibrium solutions of magnetised thick discs around a highly spinning Kerr black hole ($a = 0.99$). The study has considered non-self-gravitating, polytropic, constant angular momentum discs endowed with a purely toroidal magnetic field. The initial data have been constructed considering three different approaches. In two of them, which we labelled \texttt{KM}$h$ and \texttt{KM}$\rho$,  the magnetic field has been incorporated in a consistent way in the solution, and they differ by the fluid being relativistic or otherwise from a thermodynamical point of view. In the third approach (\texttt{MFD}) the magnetic field has been incorporated as an `ad hoc' perturbation on to an otherwise purely hydrodynamical solution. This straightforward last approach has also been adopted by previous works (e.g.~\citet{Gammie03,Noble2006,Shiokawa2012,Porth2017,Mizuta2018,Bowen2018,EHTVI}). However, those studies are based on poloidal  magnetic field setups, which lead to MRI unstable evolutions. Hence, our findings for toroidal distributions should not necessarily be taken at face value for other types of field setups. 

The initial data have been perturbed and evolved up to a final time of about 100 orbital periods using the \texttt{BHAC} code~\citep{Porth2017} which solves the non-linear GRMHD equations. We have analysed the stability properties of the initial data under a small perturbation that triggers the accretion of mass and angular momentum on to the black hole. The various outcomes of the different prescriptions used to account for the magnetic field  have been compared for increasingly larger values of the disc magnetisation. We have explored, in particular, four representative values of the magnetisation parameter $\beta_{\rm m}$ spanning from almost hydrodynamical discs to very strongly magnetised tori.

Notable differences have been found in the long-term evolutions of the initial data. Most importantly, our study has revealed that highly magnetised discs (namely, $\beta_{\rm m} = 10^{-3}$) are unstable, and hence prone to be accreted or expelled, unless the initial data incorporate the magnetic field in a self-consistent way. Only for weak magnetic fields, the long-term evolution of the models is unaffected by the way the magnetic field is incorporated in the initial data. We note, in particular, that in the simulations by the EHT Collaboration, despite the magnetic field is not consistently built in, the values of the magnetisation parameter are sufficiently small ($\beta_{\rm m} = 10^{2}$) not to artificially affect the stability of the discs. In our \textit{consistent} approaches the evolution leads to the formation of smaller \textit{mini-discs} with weaker magnetisation when compared to the initial state, surrounded by a highly magnetised, low density envelope. In general we find that  the disc angular momentum increases during the evolution with respect to the initial constant value and the discs become smaller and stripped of any external material for increasing values of the magnetisation. This is in agreement with previous results from~\cite{Wielgus2015}, who found that magnetised discs ($\beta_{\rm m} = 0.1$ , $1.0$) are stable to axisymmetric perturbation, although those simulations are fairly short, extending only $t \sim 4 t_{\rm orb}$. Our simulations are also consistent with those of~\cite{Montero07} (again, significantly shorter) where the frequencies of quasi-periodic oscillations of the discs were computed from a quasi-stable configuration for weakly and mildly magnetised discs. 

Two obvious limitations of this work have to do with our simplifying assumptions. Firstly, a constant specific angular momentum distribution is simplistic and unrealistic. And secondly, a purely toroidal magnetic field distribution is very unlikely to exist in a realistic astrophysical scenario (see~\cite{Ioka2003} for a discussion on this topic). Therefore, we could extend this study in two directions, namely i) considering non-constant angular momentum distributions, and ii) considering poloidal magnetic field distributions (to this end, we first have to construct \textit{consistent} initial data). Finally, we note that the similarities we have found in the evolutions of approaches \texttt{KM}$h$ and \texttt{KM}$\rho$ are expected due to the small deviation of the value $h \simeq 1$ for a potential gap of $|\Delta W| = 0.2216$. However, other types of compact objects might provide larger potential gaps (\eg the Kerr black holes with scalar hair described in~\cite{Gimeno-Soler:2019} achieve values of $|\Delta W| > 1$). For this reason, we could expect to find differences in the evolution between approaches \texttt{KM}$h$ and \texttt{KM}$\rho$ for such central objects even for low magnetised discs. This study will be reported elsewhere.

\section*{Acknowledgements}

We thank Oliver Porth for his comments on this work and his aid with the \texttt{BHAC} code.
ACO gratefully acknowledges support from a CONACYT Postdoctoral Fellowship (291168, 291258). 
JAF acknowledges financial support provided by the Spanish  Agencia Estatal de Investigaci\'on  (grants AYA2015-66899-C2-1-P 
and PGC2018-095984-B-I00),  by the  Generalitat  Valenciana  (PROMETEO/2019/071) and by the European Union's Horizon 2020 
research and innovation (RISE) programme H2020-MSCA-RISE-2017  (Grant  No. FunFiCO-777740).
Computations have been performed at the Lluis Vives cluster of the Universitat de Val\`encia and at the Iboga cluster of the Goethe Universit\"at Frankfurt.

\bibliographystyle{mnras}
\bibliography{aeireferences}

\begin{thebibliography}{}
\makeatletter
\relax
\def\mn@urlcharsother{\let\do\@makeother \do\$\do\&\do\#\do\^\do\_\do\%\do\~}
\def\mn@doi{\begingroup\mn@urlcharsother \@ifnextchar [ {\mn@doi@}
  {\mn@doi@[]}}
\def\mn@doi@[#1]#2{\def\@tempa{#1}\ifx\@tempa\@empty \href
  {http://dx.doi.org/#2} {doi:#2}\else \href {http://dx.doi.org/#2} {#1}\fi
  \endgroup}
\def\mn@eprint#1#2{\mn@eprint@#1:#2::\@nil}
\def\mn@eprint@arXiv#1{\href {http://arxiv.org/abs/#1} {{\tt arXiv:#1}}}
\def\mn@eprint@dblp#1{\href {http://dblp.uni-trier.de/rec/bibtex/#1.xml}
  {dblp:#1}}
\def\mn@eprint@#1:#2:#3:#4\@nil{\def\@tempa {#1}\def\@tempb {#2}\def\@tempc
  {#3}\ifx \@tempc \@empty \let \@tempc \@tempb \let \@tempb \@tempa \fi \ifx
  \@tempb \@empty \def\@tempb {arXiv}\fi \@ifundefined
  {mn@eprint@\@tempb}{\@tempb:\@tempc}{\expandafter \expandafter \csname
  mn@eprint@\@tempb\endcsname \expandafter{\@tempc}}}

\bibitem[\protect\citeauthoryear{{Abramowicz}, {Jaroszynski}  \&
  {Sikora}}{{Abramowicz} et~al.}{1978}]{Abramowicz78}
{Abramowicz} M.,  {Jaroszynski} M.,   {Sikora} M.,  1978, Astron. Astrophys.,
  \href {http://adsabs.harvard.edu/abs/1978A%26A....63..221A} {63, 221}

\bibitem[\protect\citeauthoryear{{Abramowicz}, {Calvani}  \&
  {Nobili}}{{Abramowicz} et~al.}{1983}]{Abramowicz83}
{Abramowicz} M.~A.,  {Calvani} M.,   {Nobili} L.,  1983, \mn@doi [Nature]
  {10.1038/302597a0}, \href {http://adsabs.harvard.edu/abs/1983Natur.302..597A}
  {302, 597}

\bibitem[\protect\citeauthoryear{Acker, B.~de R.~Borges  \& Costa}{Acker
  et~al.}{2016}]{Acker2016}
Acker F.,  B.~de R.~Borges R.,   Costa B.,  2016, \mn@doi [J. Comput. Phys.]
  {10.1016/j.jcp.2016.01.038}, 313, 726

\bibitem[\protect\citeauthoryear{Anninos, Fragile  \& Salmonson}{Anninos
  et~al.}{2005}]{Anninos05c}
Anninos P.,  Fragile P.~C.,   Salmonson J.~D.,  2005, \mn@doi [Astrophys. J.]
  {10.1086}, 635, 723

\bibitem[\protect\citeauthoryear{Anninos, Bryant, Fragile, Holgado, Lau  \&
  Nemergut}{Anninos et~al.}{2017}]{Anninos2017}
Anninos P.,  Bryant C.,  Fragile P.,  Holgado A.,  Lau C.,   Nemergut D.,
  2017, The Astrophysical Journal Supplement Series, 231, 17

\bibitem[\protect\citeauthoryear{Ant{\'o}n, Zanotti, Miralles, Mart{\'\i},
  Ib{\'a}{\~n}ez, Font  \& Pons}{Ant{\'o}n et~al.}{2006}]{Anton05}
Ant{\'o}n L.,  Zanotti O.,  Miralles J.~A.,  Mart{\'\i} J.~M.,  Ib{\'a}{\~n}ez
  J.~M.,  Font J.~A.,   Pons J.~A.,  2006, Astrophys. J., 637, 296

\bibitem[\protect\citeauthoryear{Baiotti \& Rezzolla}{Baiotti \&
  Rezzolla}{2017}]{Baiotti2016}
Baiotti L.,  Rezzolla L.,  2017, \mn@doi [Rept. Prog. Phys.]
  {10.1088/1361-6633/aa67bb}, 80, 096901

\bibitem[\protect\citeauthoryear{{Bowen}, {Mewes}, {Campanelli}, {Noble},
  {Krolik}  \& {Zilh{\~a}o}}{{Bowen} et~al.}{2018}]{Bowen2018}
{Bowen} D.~B.,  {Mewes} V.,  {Campanelli} M.,  {Noble} S.~C.,  {Krolik} J.~H.,
   {Zilh{\~a}o} M.,  2018, \mn@doi [\apjl] {10.3847/2041-8213/aaa756}, \href
  {http://adsabs.harvard.edu/abs/2018ApJ...853L..17B} {853, L17}

\bibitem[\protect\citeauthoryear{{Bugli}, {Guilet}, {M{\"u}ller}, {Del Zanna},
  {Bucciantini}  \& {Montero}}{{Bugli} et~al.}{2018}]{Bugli2018}
{Bugli} M.,  {Guilet} J.,  {M{\"u}ller} E.,  {Del Zanna} L.,  {Bucciantini} N.,
    {Montero} P.~J.,  2018, \mn@doi [\mnras] {10.1093/mnras/stx3158}, \href
  {http://adsabs.harvard.edu/abs/2018MNRAS.475..108B} {475, 108}

\bibitem[\protect\citeauthoryear{{Chan}, {Psaltis}, {{\"O}zel}, {Narayan}  \&
  {Sa{\c d}owski}}{{Chan} et~al.}{2015}]{Chan15}
{Chan} C.-K.,  {Psaltis} D.,  {{\"O}zel} F.,  {Narayan} R.,   {Sa{\c d}owski}
  A.,  2015, \mn@doi [Astrophysical Journal] {10.1088/0004-637X/799/1/1}, \href
  {http://adsabs.harvard.edu/abs/2015ApJ...799....1C} {799, 1}

\bibitem[\protect\citeauthoryear{{Daigne} \& {Font}}{{Daigne} \&
  {Font}}{2004}]{Daigne04}
{Daigne} F.,  {Font} J.~A.,  2004, \mn@doi [Mon. Not. R. Astron. Soc.]
  {10.1111/j.1365-2966.2004.07547.x}, \href
  {http://adsabs.harvard.edu/abs/2004MNRAS.349..841D} {349, 841}

\bibitem[\protect\citeauthoryear{{De Villiers} \& {Hawley}}{{De Villiers} \&
  {Hawley}}{2003a}]{DeVilliers03a}
{De Villiers} J.-P.,  {Hawley} J.~F.,  2003a, \mn@doi [\apj] {10.1086/373949},
  \href {http://adsabs.harvard.edu/abs/2003ApJ...589..458D} {589, 458}

\bibitem[\protect\citeauthoryear{{De Villiers} \& {Hawley}}{{De Villiers} \&
  {Hawley}}{2003b}]{DeVilliers03}
{De Villiers} J.-P.,  {Hawley} J.~F.,  2003b, \mn@doi [\apj] {10.1086/375866},
  \href {http://adsabs.harvard.edu/abs/2003ApJ...592.1060D} {592, 1060}

\bibitem[\protect\citeauthoryear{Dexter \& Fragile}{Dexter \&
  Fragile}{2011}]{Dexter2011}
Dexter J.,  Fragile P.~C.,  2011, The Astrophysical Journal, 730, 36

\bibitem[\protect\citeauthoryear{{Dexter}, {McKinney}  \& {Agol}}{{Dexter}
  et~al.}{2012}]{Dexter2012}
{Dexter} J.,  {McKinney} J.~C.,   {Agol} E.,  2012, \mn@doi [Mon. Not. R.
  Astron. Soc.] {10.1111/j.1365-2966.2012.20409.x}, \href
  {http://adsabs.harvard.edu/abs/2012MNRAS.421.1517D} {421, 1517}

\bibitem[\protect\citeauthoryear{Einfeldt}{Einfeldt}{1988}]{Einfeldt88}
Einfeldt B.,  1988, SIAM J. Numer. Anal., 25, 294

\bibitem[\protect\citeauthoryear{{Event Horizon Telescope Collaboration}
  et~al.,}{{Event Horizon Telescope Collaboration} et~al.}{2019a}]{EHTI}
{Event Horizon Telescope Collaboration} et~al., 2019a, \mn@doi [\apjl]
  {10.3847/2041-8213/ab0ec7}, \href
  {https://ui.adsabs.harvard.edu/abs/2019ApJ...875L...1E} {875, L1}

\bibitem[\protect\citeauthoryear{{Event Horizon Telescope Collaboration}
  et~al.,}{{Event Horizon Telescope Collaboration} et~al.}{2019b}]{EHTVI}
{Event Horizon Telescope Collaboration} et~al., 2019b, \mn@doi [\apjl]
  {10.3847/2041-8213/ab1141}, \href
  {https://ui.adsabs.harvard.edu/abs/2019ApJ...875L...6E} {875, L6}

\bibitem[\protect\citeauthoryear{{Fishbone} \& {Moncrief}}{{Fishbone} \&
  {Moncrief}}{1976}]{Fishbone76}
{Fishbone} L.~G.,  {Moncrief} V.,  1976, Astrophys. J., \href
  {http://adsabs.harvard.edu/abs/1976ApJ...207..962F} {207, 962}

\bibitem[\protect\citeauthoryear{{Font} \& {Daigne}}{{Font} \&
  {Daigne}}{2002}]{Font02a}
{Font} J.~A.,  {Daigne} F.,  2002, \mn@doi [Mon. Not. R. Astron. Soc.]
  {10.1046/j.1365-8711.2002.05515.x}, \href
  {http://adsabs.harvard.edu/abs/2002MNRAS.334..383F} {334, 383}

\bibitem[\protect\citeauthoryear{{Foucart}, {Chandra}, {Gammie}  \&
  {Quataert}}{{Foucart} et~al.}{2016}]{Foucart2015b}
{Foucart} F.,  {Chandra} M.,  {Gammie} C.~F.,   {Quataert} E.,  2016, \mn@doi
  [Mon. Not. R. Astron. Soc.] {10.1093/mnras/stv2687}, \href
  {http://adsabs.harvard.edu/abs/2016MNRAS.456.1332F} {456, 1332}

\bibitem[\protect\citeauthoryear{{Fragile} \& {S{\c a}dowski}}{{Fragile} \&
  {S{\c a}dowski}}{2017}]{Fragile2017}
{Fragile} P.~C.,  {S{\c a}dowski} A.,  2017, \mn@doi [Mon. Not. R. Astron.
  Soc.] {10.1093/mnras/stx274}, 467, 1838

\bibitem[\protect\citeauthoryear{Fragile, Blaes, Anninois  \&
  Salmonson}{Fragile et~al.}{2007}]{Fragile:2007dk}
Fragile P.~C.,  Blaes O.~M.,  Anninois P.,   Salmonson J.~D.,  2007, \mn@doi
  [Astrophys J.] {10.1086/521092}, \href
  {http://adsabs.harvard.edu/abs/2007ApJ...668..417F} {668, 417}

\bibitem[\protect\citeauthoryear{Gammie, McKinney  \& T{\'o}th}{Gammie
  et~al.}{2003}]{Gammie03}
Gammie C.~F.,  McKinney J.~C.,   T{\'o}th G.,  2003, Astrophys. J., 589, 458

\bibitem[\protect\citeauthoryear{{Gimeno-Soler} \& {Font}}{{Gimeno-Soler} \&
  {Font}}{2017}]{Gimeno-Soler:2017}
{Gimeno-Soler} S.,  {Font} J.~A.,  2017, \mn@doi [\aap]
  {10.1051/0004-6361/201730935}, \href
  {https://ui.adsabs.harvard.edu/#abs/2017A&A...607A..68G} {607, A68}

\bibitem[\protect\citeauthoryear{{Gimeno-Soler}, {Font}, {Herdeiro}  \&
  {Radu}}{{Gimeno-Soler} et~al.}{2019}]{Gimeno-Soler:2019}
{Gimeno-Soler} S.,  {Font} J.~A.,  {Herdeiro} C.,   {Radu} E.,  2019, \mn@doi
  [\prd] {10.1103/PhysRevD.99.043002}, \href
  {https://ui.adsabs.harvard.edu/abs/2019PhRvD..99d3002G} {99, 043002}

\bibitem[\protect\citeauthoryear{Harten, Lax  \& van Leer}{Harten
  et~al.}{1983}]{Harten83}
Harten A.,  Lax P.~D.,   van Leer B.,  1983, \mn@doi [SIAM Rev.]
  {10.1137/1025002}, 25, 35

\bibitem[\protect\citeauthoryear{{Hawley}, {Guan}  \& {Krolik}}{{Hawley}
  et~al.}{2011}]{Hawley2011}
{Hawley} J.~F.,  {Guan} X.,   {Krolik} J.~H.,  2011, \mn@doi [\apj]
  {10.1088/0004-637X/738/1/84}, \href
  {https://ui.adsabs.harvard.edu/#abs/2011ApJ...738...84H} {738, 84}

\bibitem[\protect\citeauthoryear{{Ioka} \& {Sasaki}}{{Ioka} \&
  {Sasaki}}{2003}]{Ioka2003}
{Ioka} K.,  {Sasaki} M.,  2003, \mn@doi [Phys. Rev. D]
  {10.1103/PhysRevD.67.124026}, \href
  {http://adsabs.harvard.edu/abs/2003PhRvD..67l4026I} {67, 124026}

\bibitem[\protect\citeauthoryear{{Janiuk}, {Sapountzis}, {Mortier}  \&
  {Janiuk}}{{Janiuk} et~al.}{2018}]{Janiuk2018}
{Janiuk} A.,  {Sapountzis} K.,  {Mortier} J.,   {Janiuk} I.,  2018, preprint,
  \href {http://adsabs.harvard.edu/abs/2018arXiv180511305J} {} (\mn@eprint
  {arXiv} {1805.11305})

\bibitem[\protect\citeauthoryear{{Koide}, {Shibata}  \& {Kudoh}}{{Koide}
  et~al.}{1999}]{Koide99}
{Koide} S.,  {Shibata} K.,   {Kudoh} T.,  1999, \mn@doi [Astrophys. J.]
  {10.1086/307667}, \href {http://adsabs.harvard.edu/abs/1999ApJ...522..727K}
  {522, 727}

\bibitem[\protect\citeauthoryear{{Komissarov}}{{Komissarov}}{2006}]{Komissarov2006a}
{Komissarov} S.~S.,  2006, \mn@doi [Mon. Not. R. Astron. Soc.]
  {10.1111/j.1365-2966.2006.10183.x}, \href
  {http://adsabs.harvard.edu/abs/2006MNRAS.368..993K} {368, 993}

\bibitem[\protect\citeauthoryear{{Korobkin}, {Abdikamalov}, {Schnetter},
  {Stergioulas}  \& {Zink}}{{Korobkin} et~al.}{2011}]{Korobkin2011}
{Korobkin} O.,  {Abdikamalov} E.~B.,  {Schnetter} E.,  {Stergioulas} N.,
  {Zink} B.,  2011, \mn@doi [Phys. Rev. D] {10.1103/PhysRevD.83.043007}, \href
  {http://adsabs.harvard.edu/abs/2011PhRvD..83d3007K} {83, 043007}

\bibitem[\protect\citeauthoryear{{L\"{o}hner}}{{L\"{o}hner}}{1987}]{Loehner87}
{L\"{o}hner} R.,  1987, \mn@doi [Computer Methods in Applied Mechanics and
  Engineering] {10.1016/0045-7825(87)90098-3}, \href
  {http://cdsads.u-strasbg.fr/abs/1987CMAME..61..323L} {61, 323}

\bibitem[\protect\citeauthoryear{{Mach}, {Gimeno-Soler}, {Font},
  {Odrzywo{\l}ek}  \& {Pir{\'o}g}}{{Mach} et~al.}{2019}]{Mach:2019}
{Mach} P.,  {Gimeno-Soler} S.,  {Font} J.~A.,  {Odrzywo{\l}ek} A.,
  {Pir{\'o}g} M.,  2019, \mn@doi [\prd] {10.1103/PhysRevD.99.104063}, \href
  {https://ui.adsabs.harvard.edu/abs/2019PhRvD..99j4063M} {99, 104063}

\bibitem[\protect\citeauthoryear{{Matsumoto}, {Miyoshi}  \&
  {Takasao}}{{Matsumoto} et~al.}{2019}]{Matsumoto2019}
{Matsumoto} T.,  {Miyoshi} T.,   {Takasao} S.,  2019, \mn@doi [\apj]
  {10.3847/1538-4357/ab05cb}, \href
  {https://ui.adsabs.harvard.edu/abs/2019ApJ...874...37M} {874, 37}

\bibitem[\protect\citeauthoryear{{McKinney} \& {Blandford}}{{McKinney} \&
  {Blandford}}{2009}]{McKinney2009}
{McKinney} J.~C.,  {Blandford} R.~D.,  2009, \mn@doi [Mon. Not. R. Astron.
  Soc.] {10.1111/j.1745-3933.2009.00625.x}, \href
  {http://adsabs.harvard.edu/abs/2009MNRAS.394L.126M} {394, L126}

\bibitem[\protect\citeauthoryear{{McKinney}, {Tchekhovskoy}  \&
  {Blandford}}{{McKinney} et~al.}{2012}]{McKinney2012}
{McKinney} J.~C.,  {Tchekhovskoy} A.,   {Blandford} R.~D.,  2012, \mn@doi [Mon.
  Not. R. Astron. Soc.] {10.1111/j.1365-2966.2012.21074.x}, \href
  {http://adsabs.harvard.edu/abs/2012MNRAS.423.3083M} {423, 3083}

\bibitem[\protect\citeauthoryear{{McKinney}, {Tchekhovskoy}, {Sadowski}  \&
  {Narayan}}{{McKinney} et~al.}{2014}]{McKinney2014}
{McKinney} J.~C.,  {Tchekhovskoy} A.,  {Sadowski} A.,   {Narayan} R.,  2014,
  \mn@doi [Mon. Not. R. Astron. Soc.] {10.1093/mnras/stu762}, \href
  {http://cdsads.u-strasbg.fr/abs/2014MNRAS.441.3177M} {441, 3177}

\bibitem[\protect\citeauthoryear{{Mewes}, {Galeazzi}, {Font}, {Montero}  \&
  {Stergioulas}}{{Mewes} et~al.}{2016}]{Mewes2016}
{Mewes} V.,  {Galeazzi} F.,  {Font} J.~A.,  {Montero} P.~J.,   {Stergioulas}
  N.,  2016, \mn@doi [Mon. Not. R. Astron. Soc.] {10.1093/mnras/stw1490}, \href
  {http://adsabs.harvard.edu/abs/2016MNRAS.461.2480M} {461, 2480}

\bibitem[\protect\citeauthoryear{Mignone, Ugliano  \& Bodo}{Mignone
  et~al.}{2009}]{Mignone2009}
Mignone a.,  Ugliano M.,   Bodo G.,  2009, \mn@doi [Mon. Not. R. Astron. Soc.]
  {10.1111/j.1365-2966.2008.14221.x}, 393, 1141

\bibitem[\protect\citeauthoryear{{Mizuta}, {Ebisuzaki}, {Tajima}  \&
  {Nagataki}}{{Mizuta} et~al.}{2018}]{Mizuta2018}
{Mizuta} A.,  {Ebisuzaki} T.,  {Tajima} T.,   {Nagataki} S.,  2018, \mn@doi
  [\mnras] {10.1093/mnras/sty1453}, \href
  {http://adsabs.harvard.edu/abs/2018MNRAS.479.2534M} {479, 2534}

\bibitem[\protect\citeauthoryear{{Montero}, {Zanotti}, {Font}  \&
  {Rezzolla}}{{Montero} et~al.}{2007}]{Montero07}
{Montero} P.~J.,  {Zanotti} O.,  {Font} J.~A.,   {Rezzolla} L.,  2007, \mn@doi
  [Mon. Not. R. Astron. Soc.] {10.1111/j.1365-2966.2007.11844.x}, \href
  {http://adsabs.harvard.edu/abs/2007MNRAS.378.1101M} {378, 1101}

\bibitem[\protect\citeauthoryear{{Noble}, {Gammie}, {McKinney}  \& {Del
  Zanna}}{{Noble} et~al.}{2006}]{Noble2006}
{Noble} S.~C.,  {Gammie} C.~F.,  {McKinney} J.~C.,   {Del Zanna} L.,  2006,
  \mn@doi [Astrophys. J.] {10.1086/500349}, \href
  {http://adsabs.harvard.edu/abs/2006ApJ...641..626N} {641, 626}

\bibitem[\protect\citeauthoryear{Olivares~S\'anchez, Porth  \&
  Mizuno}{Olivares~S\'anchez et~al.}{2018}]{Olivares2018b}
Olivares~S\'anchez H.,  Porth O.,   Mizuno Y.,  2018, \mn@doi [J. Phys. Conf.
  Ser.] {10.1088/1742-6596/1031/1/012008}, 1031, 012008

\bibitem[\protect\citeauthoryear{{Papaloizou} \& {Pringle}}{{Papaloizou} \&
  {Pringle}}{1984}]{Papaloizou84}
{Papaloizou} J.~C.~B.,  {Pringle} J.~E.,  1984, Mon. Not. R. Astron. Soc.,
  \href
  {http://adsabs.harvard.edu/cgi-bin/nph-bib_query?bibcode=1984MNRAS.208..721P&db_key=AST}
  {208, 721}

\bibitem[\protect\citeauthoryear{{Penna}, {Kulkarni}  \& {Narayan}}{{Penna}
  et~al.}{2013}]{Penna2013b}
{Penna} R.~F.,  {Kulkarni} A.,   {Narayan} R.,  2013, \mn@doi [Astron.
  Astrophys.] {10.1051/0004-6361/201219666}, \href
  {http://adsabs.harvard.edu/abs/2013A%26A...559A.116P} {559, A116}

\bibitem[\protect\citeauthoryear{{Pimentel}, {Lora-Clavijo}  \&
  {Gonzalez}}{{Pimentel} et~al.}{2018a}]{Pimentel2018a}
{Pimentel} O.~M.,  {Lora-Clavijo} F.~D.,   {Gonzalez} G.~A.,  2018a, \mn@doi
  [\aap] {10.1051/0004-6361/201833736}, \href
  {http://adsabs.harvard.edu/abs/2018A%26A...619A..57P} {619, A57}

\bibitem[\protect\citeauthoryear{{Pimentel}, {Lora-Clavijo}  \&
  {Gonz{\'a}lez}}{{Pimentel} et~al.}{2018b}]{Pimentel2018b}
{Pimentel} O.~M.,  {Lora-Clavijo} F.~D.,   {Gonz{\'a}lez} G.~A.,  2018b,
  \mn@doi [\apj] {10.3847/1538-4357/aac6d0}, \href
  {http://adsabs.harvard.edu/abs/2018ApJ...861..115P} {861, 115}

\bibitem[\protect\citeauthoryear{{Porth}, {Olivares}, {Mizuno}, {Younsi},
  {Rezzolla}, {Moscibrodzka}, {Falcke}  \& {Kramer}}{{Porth}
  et~al.}{2017}]{Porth2017}
{Porth} O.,  {Olivares} H.,  {Mizuno} Y.,  {Younsi} Z.,  {Rezzolla} L.,
  {Moscibrodzka} M.,  {Falcke} H.,   {Kramer} M.,  2017, \mn@doi [Computational
  Astrophysics and Cosmology] {10.1186/s40668-017-0020-2}, 4, 1

\bibitem[\protect\citeauthoryear{Shibata}{Shibata}{2007}]{Shibata2007}
Shibata M.,  2007, \mn@doi [Phys. Rev. D] {10.1103/PhysRevD.76.064035}, 76,
  064035

\bibitem[\protect\citeauthoryear{{Shibata} \& {Sekiguchi}}{{Shibata} \&
  {Sekiguchi}}{2012}]{Shibata2012}
{Shibata} M.,  {Sekiguchi} Y.,  2012, Progress of Theoretical Physics, \href
  {http://adsabs.harvard.edu/abs/2012PThPh.127..535S} {127, 535}

\bibitem[\protect\citeauthoryear{Shibata \& Taniguchi}{Shibata \&
  Taniguchi}{2011}]{ShibataTaniguchilrr-2011-6}
Shibata M.,  Taniguchi K.,  2011, Living Rev. Relativity, 14

\bibitem[\protect\citeauthoryear{{Shiokawa}, {Dolence}, {Gammie}  \&
  {Noble}}{{Shiokawa} et~al.}{2012}]{Shiokawa2012}
{Shiokawa} H.,  {Dolence} J.~C.,  {Gammie} C.~F.,   {Noble} S.~C.,  2012,
  \mn@doi [Astrophys. J.] {10.1088/0004-637X/744/2/187}, \href
  {http://adsabs.harvard.edu/abs/2012ApJ...744..187S} {744, 187}

\bibitem[\protect\citeauthoryear{Shu \& Osher}{Shu \& Osher}{1988}]{Shu88}
Shu C.~W.,  Osher S.~J.,  1988, J. Comput. Phys., 77, 439

\bibitem[\protect\citeauthoryear{{Siegel} \& {Metzger}}{{Siegel} \&
  {Metzger}}{2017}]{Siegel2017}
{Siegel} D.~M.,  {Metzger} B.~D.,  2017, \mn@doi [Physical Review Letters]
  {10.1103/PhysRevLett.119.231102}, \href
  {http://adsabs.harvard.edu/abs/2017PhRvL.119w1102S} {119, 231102}

\bibitem[\protect\citeauthoryear{{Siegel} \& {Metzger}}{{Siegel} \&
  {Metzger}}{2018}]{Siegel2018}
{Siegel} D.~M.,  {Metzger} B.~D.,  2018, \mn@doi [\apj]
  {10.3847/1538-4357/aabaec}, \href
  {http://adsabs.harvard.edu/abs/2018ApJ...858...52S} {858, 52}

\bibitem[\protect\citeauthoryear{{Sorathia}, {Krolik}  \& {Hawley}}{{Sorathia}
  et~al.}{2013}]{Sorathia2013}
{Sorathia} K.~A.,  {Krolik} J.~H.,   {Hawley} J.~F.,  2013, \mn@doi [\apj]
  {10.1088/0004-637X/777/1/21}, \href
  {http://adsabs.harvard.edu/abs/2013ApJ...777...21S} {777, 21}

\bibitem[\protect\citeauthoryear{{Stergioulas}}{{Stergioulas}}{2011}]{Stergioulas2011}
{Stergioulas} N.,  2011, \mn@doi [International Journal of Modern Physics D]
  {10.1142/S021827181101944X}, \href
  {http://adsabs.harvard.edu/abs/2011IJMPD..20.1251S} {20, 1251}

\bibitem[\protect\citeauthoryear{{Wielgus}, {Fragile}, {Wang}  \&
  {Wilson}}{{Wielgus} et~al.}{2015}]{Wielgus2015}
{Wielgus} M.,  {Fragile} P.~C.,  {Wang} Z.,   {Wilson} J.,  2015, \mn@doi
  [\mnras] {10.1093/mnras/stu2676}, \href
  {http://adsabs.harvard.edu/abs/2015MNRAS.447.3593W} {447, 3593}

\bibitem[\protect\citeauthoryear{{Witzany} \& {Jefremov}}{{Witzany} \&
  {Jefremov}}{2018}]{Witzany2018}
{Witzany} V.,  {Jefremov} P.,  2018, \mn@doi [\aap]
  {10.1051/0004-6361/201732361}, \href
  {http://adsabs.harvard.edu/abs/2018A%26A...614A..75W} {614, A75}

\bibitem[\protect\citeauthoryear{{Zanotti} \& {Pugliese}}{{Zanotti} \&
  {Pugliese}}{2015}]{Zanotti2015d}
{Zanotti} O.,  {Pugliese} D.,  2015, \mn@doi [General Relativity and
  Gravitation] {10.1007/s10714-015-1886-4}, \href
  {https://ui.adsabs.harvard.edu/#abs/2015GReGr..47...44Z} {47, 44}

\bibitem[\protect\citeauthoryear{{von Zeipel}}{{von
  Zeipel}}{1924}]{vonZeipel1924}
{von Zeipel} H.,  1924, Mon. Not. Roy. Soc., \href
  {http://adsabs.harvard.edu/abs/1924MNRAS..84..665V} {84, 665}

\makeatother
\end{thebibliography}

\end{document}